\documentclass{aastex631} 
\usepackage{amssymb}
\usepackage{epsfig}
\usepackage{mathtools}
\usepackage{comment}
\usepackage{bm}
\usepackage{epstopdf}
\usepackage{amssymb}
\usepackage{array}
\usepackage{geometry}
\usepackage[figuresright]{rotating}
\newcolumntype{H}{>{\setbox0=\hbox\bgroup}c<{\egroup}@{}}
\usepackage[caption=false]{subfig}
\renewcommand{\baselinestretch}{1.3}
\newcommand{\kms}{\mbox{km~s$^{-1}$}}
\newcommand{\s}{\mbox{$''$}}

\newcommand{\my}{\mbox{$M_{\odot}$~yr$^{-1}$}}
\newcommand{\ls}{\mbox{$L_{\odot}$}}
\newcommand{\ms}{\mbox{$M_{\odot}$}}
\newcommand\mdot{$\dot{M}  $}

\newcommand{\fluxu}{\mbox{erg\,s$^{-1}$\,cm$^{-2}$\,\AA$^{-1}$}}

\newcommand{\intunit}    {erg\,s$^{-1}$\,cm$^{-2}$\,\AA$^{-1}$\,arcsec$^{-2}$}

\begin{document}
\title{Faint but not forgotten. I. First results from a search for astrospheres around AGB stars in the far-ultraviolet}
\author{Raghvendra Sahai}
\affiliation{Jet Propulsion Laboratory, MS 183-900, California Institute of Technology, Pasadena, CA 91109}
\author{Benjamin Stenger}
\affiliation{California State University, Fullerton, CA 92831}
\affiliation{Jet Propulsion Laboratory, MS 183-900, California Institute of Technology, Pasadena, CA 91109}
\email{raghvendra.sahai@jpl.nasa.gov}

\begin{abstract}
Using the GALEX archive, we have discovered extended structures around 10 AGB stars (out of a total 92 searched) emitting in the far-ultraviolet (FUV) band. In all but one, we find the typical morphology expected for a spherical wind moving relative to, and interacting with the ISM to produce an astrosphere. The exception is V\,Hya whose mass-ejection is known to be highly aspherical, where we find evidence of its large parabolic outflows interacting with the ISM, and its collimated, extreme velocity outflows interacting with the circumstellar medium. For 8 objects with relatively large proper motions, we find (as expected) that the termination-shock region lies in a hemisphere that contains the proper motion vector. Radial intensity cuts for each source have been used to locate the termination shock and the astropause's outer edge. In a few objects, the cuts also reveal faint emission just outside the astropause that likely arises in shocked ISM material. We have used these data, together with published mass-loss rates and wind expansion velocities, to determine the total mass lost and duration for each source -- we find that the duration of and total mass in the shocked wind are significantly larger than their corresponding values for the unshocked wind. The combination of FUV and far-IR data on AGB astrospheres, provides a unique database for theoretical studies (numerical simulations) of wind-ISM interactions. We show that a Cyclical Spatial Heterodyne Spectrometer on a small space-based telescope, can provide high-resolution spectra of astrospheres to confirm the emission mechanism.
\end{abstract}


\section{Introduction}
Most stars have winds, more or less throughout their active lives (i.e., while nuclear burning is still ongoing at their centers). The mass-loss rates and expansion speeds vary as a function of evolutionary phase and stellar mass. For example, for low-mass main-sequence (MS) stars like the Sun, the mass-loss rate is $\sim10^{-14}$\my, whereas for high-mass MS stars, e.g., OB stars, the mass-loss rates are $\sim10^{-6}-10^{-5}$\my~(O-stars: e.g., \cite{Smith2014}) and $\sim10^{-9}$\my~(B-stars: e.g., \cite{Krticka2014}). Evolved low- and intermediate- mass stars lose mass at rates of $\sim10^{-7}-10^{-5}$\my~range (e.g., \cite{Olofsson2008}), but rates can reach as high as $>10^{-4}$\my~in some objects. 

The mass-loss from stars with MS masses in the $1-8$\,\ms~range generally peaks during the Asymptotic Giant Branch (AGB) evolutionary phase, when the stars are very cool ($T_{eff}\lesssim3500$\,K), luminous ($L\sim$10,000\,\ls), and undergoing strong radial pulsations. At this stage, these stars consist of a central C+O degenerate core, surrounded by He and H-shells which undergo nuclear-burning, and a very large stellar envelope. The heavy mass-loss, believed to be driven by radiation pressure on dust grains that condense in the cool material levitated above the photosphere as a result of pulsations, produces a dusty, molecular (H$_2$), spherical, expanding (expansion speed, $V_{e}\sim5-20$\,\kms), circumstellar envelope (CSE) around the AGB star. The mass-loss from these stars enriches the ISM with products of nucleosynthesis (incuding the biogenic elements C and N), as well as dust grains, which play a crucial role in the formation of solar systems and planets. The mass-loss rate history of these stars determines the course of their late evolution and final demise. 

However, it has been a major observational challenge to trace the full history of heavy mass-loss in these stars. The standard tracers of the mass-loss history from ground-based observations are (i) millimeter-wave CO line emission from gas, and (ii) scattered light from dust in the CSE. Observations of atomic hydrogen (HI) from the wind, generally resulting from the photdissociation of H$_2$ in the molecular wind, are usually strongly confused by Galactic emission, although it has been possible to detect HI emission for a few stars (e.g., \citep{Matthews2008,Matthews2013,Matthews2015}). CO line-emission becomes undetectable, due to photodissociation by the interstellar UV field, at radii typically $\lesssim 2\times10^{17}$\,cm (or less) even for mass-loss rates as high as $\sim10^{-5}$\,\my~(e.g., \cite{Saberi2019,Ramstedt2020}), corresponding to a mass-ejection time-scale of about 6,500\,yr for a typical expansion velocity of 10\,\kms. Dust scattered light becomes undetectable due to sensitivity at a comparable radius for smilar mass-loss rates, and the timescales probed are similar (e.g., \cite{Mauron2006,Mauron2013}). Hence, both the progenitor mass, and the total amount of mass ejected into the ISM, M$_{ejecta}$ (which depends on the envelope's outer extent -- e.g. M$_{ejecta}\propto r_{out}$, for a constant mass-loss rate at a constant expansion velocity) remains unknown. For example, in the case of IRC+10216, the best-studied mass-losing AGB star, the CSE seen in the above tracers extends to about $200{''}$ ($3.5\times10^{17}$cm), and the inferred ejecta mass is only M$_{ejecta}\sim0.15$\,\ms, a small fraction of what this star has had to have lost, given its late evolutionary phase.

The unexpected discovery of a bow-shock structure and a turbulent wake extending over 2\arcdeg~($\sim$4\,pc) in the sky, towards the AGB star, Mira, in a GALEX far-ultraviolet (FUV) image \citep{Martin2007}, resulted in a new method of probing the mass-loss history in AGB stars on very long time-scales compared to the standard probes. Following the above discovery, very extended shock structures were also found around the carbon stars IRC+10216 and CIT\,6 in GALEX FUV images (\cite{Sahai2010,Sahai2014}), who concluded that these structures result from the interaction of these stars' molecular winds with the interstellar medium (ISM), as they move through the latter. A far-infrared survey of a sample of 78 evolved stars (AGB stars and red supergiants), using the PACS instrument onboard the Herschel Space Observatory revealed bow-shocks resulting from wind-ISM interactions in $\sim40$\% of the sample \citep{Cox2012} (some of the objects in this study had been known previously to show similar far-IR emission, e.g., R\,Hya \citep{Ueta2006}, $\alpha$Ori \citep{Ueta2008b,Decin2012}, R\,Cas (e.g., \cite{Ueta2010}), and U\,Hya \citep{Izumiura2011}. 

Surveys using all-sky archives have been used extensively to catalog and study bow-shocks around massive stars (e.g., \cite{Peri2015,Kobulnicky2017,Brown2005}). We are therefore carrying out a survey of GALEX images of a large sample of AGB stars, with the primary goals of searching for and characterising extended circumstellar structures around these objects, and using these to investigate their mass-loss history over unprecedented long time-scales, and its implications for their evolutionary status.

In this paper, we present first results from a survey of objects for which long-exposure GALEX images are available and have been examined so far. Future papers will 
focus on results from surveying the full GALEX archive of images. From our current survey, we have found 10 objects with extended UV emission\footnote{this number excludes 3 objects detected previously -- Mira, IRC+10216 and CIT\,6}. We do not discuss emission from the central stars in this paper. The plan of our paper is as follows. {We first provide a summary of the archival data that we analysed, and the methodology used to search for the presence of extended UV emission associated with AGB stars (\S\,\ref{obs}). In \S\,\ref{struct}, we present our observational results for each object in which we found extended UV emission, together with the analysis used to characterise this emission. In \S\,\ref{astros}, we quantitatively analyze the wind-ISM interaction and discuss the implications of our results for the mass-loss histories of these objects. In \S\,\ref{shs}, we present a model spectrum of the FUV emission from an astrosphere assuming the current hypothesis for its origin, together with reference to an instrumental concept that can carry out the spectroscopic observations required to test this hypothesis. Finally, in \S\,\ref{sumry}, we present the main conclusions of our study.} 

\section{Archival UV Observations}\label{obs}
In our survey, we have first focused our attention on 92 AGB stars that lie in the fields-of-view of long-exposure ($>$700\,s) FUV GALEX images \citep{Morrissey2005}, generally taken as part of the Medium Imaging Survey (MIS) and various Guest Investigator Programs (GI). The GALEX archive (https://galex.stsci.edu) contains FUV and near-UV (NUV) images with a bandpass (angular resolution) of 1344-1786\,\AA\,($4.5{''}$) and 1771-2831\,\AA\,($6.0{''}$), respectively, with a pixel size of $1.5{''}\times1.5{''}$ and a field-of-view (FOV) of $1\arcdeg.25$. A total of 92 such objects were found. The associated FUV and NUV images were downloaded from the archive, together with the associated GALEX point-source FUV and NUV catalogs for these fields. 

\subsection{Image Analysis}
We followed a similar methodology as in \cite{Sahai2014} to search for faint UV emission. First, all point sources listed in the UV point-source catalogs were removed using a customised IDL routine which replaces a small region covering each star's PSF with a tile of random noise representative of the surrounding sky. The sky noise was sampled separately at the four corners of each tile and linearly interpolated throughout, so as to preserve gradients in the local sky background to first order. In some situations the field stars are not properly removed, this can occur if there are many stars close together or if the star is very bright, in either case a residue is left on the image. Two other artifacts that appear on the UV images are ghosts and hotspots. Both artifacts appear as doughnut shapes and can be clearly identified with the original image.  
After removing the point sources, the images are smoothed using IRAF's Gaussian smoothing function, ``gauss", with a FWHM of 5 pixels ($7\farcs5$), except for for U\,Ant, where we used 3 pixels ($4\farcs5$) as the emission is relatively bright.

Since the emission is relatively faint in most cases, and  emission associated with interstellar ``cirrus" is generally also present in the full circular field-of-view, we used the following criteria to identify UV emission associated with our targets.

\begin{enumerate}
\item There is extended UV emission all around the star that peaks in some part of a reasonably well-defined geometrical structure -- the latter may be (a) circularly symmetric (ring), or have either (b) a fan-shaped morphology or (c) a head-tail morphology, or (d) a combination of $b$ and $c$.
\item There are extended regions of significantly lower-level (compared to the above) intensity around the above structure that can be identified as the general ISM.
\item For each source with non-circular emission morphology, we looked for a rough axis of symmetry for the structure, using the following strategy. The indicators we focused on were an increased brightness along part of the structure and a (roughly) diametrically-opposed fainter tail. If the structure had a tail we started our search directly across from the tail, for example if the tail was aligned along, say, $PA=0^{\circ}$ we looked for a brightening along $PA=180^{\circ}$. If there was no easily identifiable tail structure, then we looked for a relatively brighter section on the periphery of the structure.

We then made radial intensity cuts averaged over an azimuthal wedge with its apex centered on the star spanning a wide range in position-angles around the symmetry axis. We generally used a fairly wide opening angle for the wedge, $\sim80\arcdeg$ in order to reduce sensitivity to local bright spots in the FUV emission, since the FUV emission is  quite faint and noisy. For each cut, we visually traced the intensity from large radii (where the cut intensity is equal to the average sky backgroud intensity) to smaller radii, and looked for a steep rise of the intensity  expected in the region of interaction of the stellar with the ambient medium. We checked that this intensity rise is not sensitive to the opening angle of the azimuthal wedge, by inspecting cuts with a range of opening angles ($\sim40\arcdeg-70\arcdeg$). 

\end{enumerate}

The analysis of such radial intensity cuts for the astrospheres of IRC\,10216 and CIT\,6 shows that the termination shock is located at the peak of the intensity rise (R$_1$, Fig.\,3 in \cite{Sahai2010}), and the thickness of the astrosheath\footnote{see Fig.\,2d in \cite{Ueta2008a} for a definition of the terms termination shock, astrosheath, astropause, bow-shock used to describe an astrosphere} is given by the distance between the peak and the radius at the which the steeply falling intensity levels off (R$_c$, Fig.\,3 in \cite{Sahai2010}), either to a low-intensity plateau region that is brighter than the average sky background, or to the average sky background. The low-intensity plateau region likely represents emission from swept-up ISM material between the outer edge of the astropause and the bow-shock interface separating the shocked and unshocked ISM (R$_2$, Fig.\,3 in \cite{Sahai2010}).

Using the above methodology, we found extended FUV structures around 10 stars. A log of the image fits files for these sources, together with the exposure times, are given in Table\,\ref{tbl:log}. We discuss these below, in order of increasing right ascension. One object, VX\,Eri, for which the association is tentative, is discussed at the end.  In Table\,\ref{tbl:mdotetc}, we list specific published stellar properties (name (Col.\,1)), Galactic coordinates (Cols.\,2--3), proper motion (from $GAIA$ DR3, \cite{Gaia2022}, Col.\,4)\footnote{with the exception of W\,Hya, where we provide the Hipparcos value from \cite{vanLeeuwen2007}, since it is not listed in $GAIA$ DR3} and mass-loss properties derived from CO data and modeling  (mass-loss rate (Col.\,6), radial velocity (Col.\,7) wind expansion velocity (Col.\,8)), and the assumed CO-to-H$_2$ abundance ratio (Col.\,9) and adopted distance for the mass-loss estimation (Col.\,5)). We also list observed properties of 
the extended FUV emission related to the wind-ISM interaction, which include an estimate of the average FUV intensity in the interaction region (Col.\,11), the termination shock radius (R$_1$) and the outer radius of the astropause (R$_c$) extracted from the radial intensity cuts (Cols.\,12--13), and characteristics of the emission morphology (Col.\,15). Conservative estimates of the errors are (i) R$_1$: $\lesssim5$\,\%, when a sharp peak is seen at the termination shock, otherwise the error is typically $\lesssim10$\,\%, (ii) R$_c$: $\lesssim10$\,\%, and (iii) R$_2$: $\lesssim15$\,\%.

For VX\,Eri no published CO data could be found, hence we have used its IRAS 60\,\micron~flux (0.62\,Jy) to derive a rough estimate of the dust-mass loss rate, \mdot(d), using the methodology given by \cite{Jura1986}. The bolometric flux, $F=1.55\times10^{-7}$\,erg\,cm$^{-2}$\,s$^{-1}$, and the mean wavelength for emission, $\lambda_e=1.7$\,\micron~required for this method, were estimated from the spectral-energy-distribution (SED), extracted using the VizieR Photometry viewer (http://vizier.cds.unistra.fr/vizier/sed/) over the $0.35-60$\,\micron~range. The gas mass-loss rate was derived from \mdot(d), assuming a typical gas-to-dust ratio of 200; we also assume a typical circumstellar expansion velocity of 10\,\kms~for this object.

\section{Results}\label{structure}\label{struct}

\subsection{EY\,Hya}
\begin{figure}[htbp]
\begin{center}
\includegraphics[width=16cm]{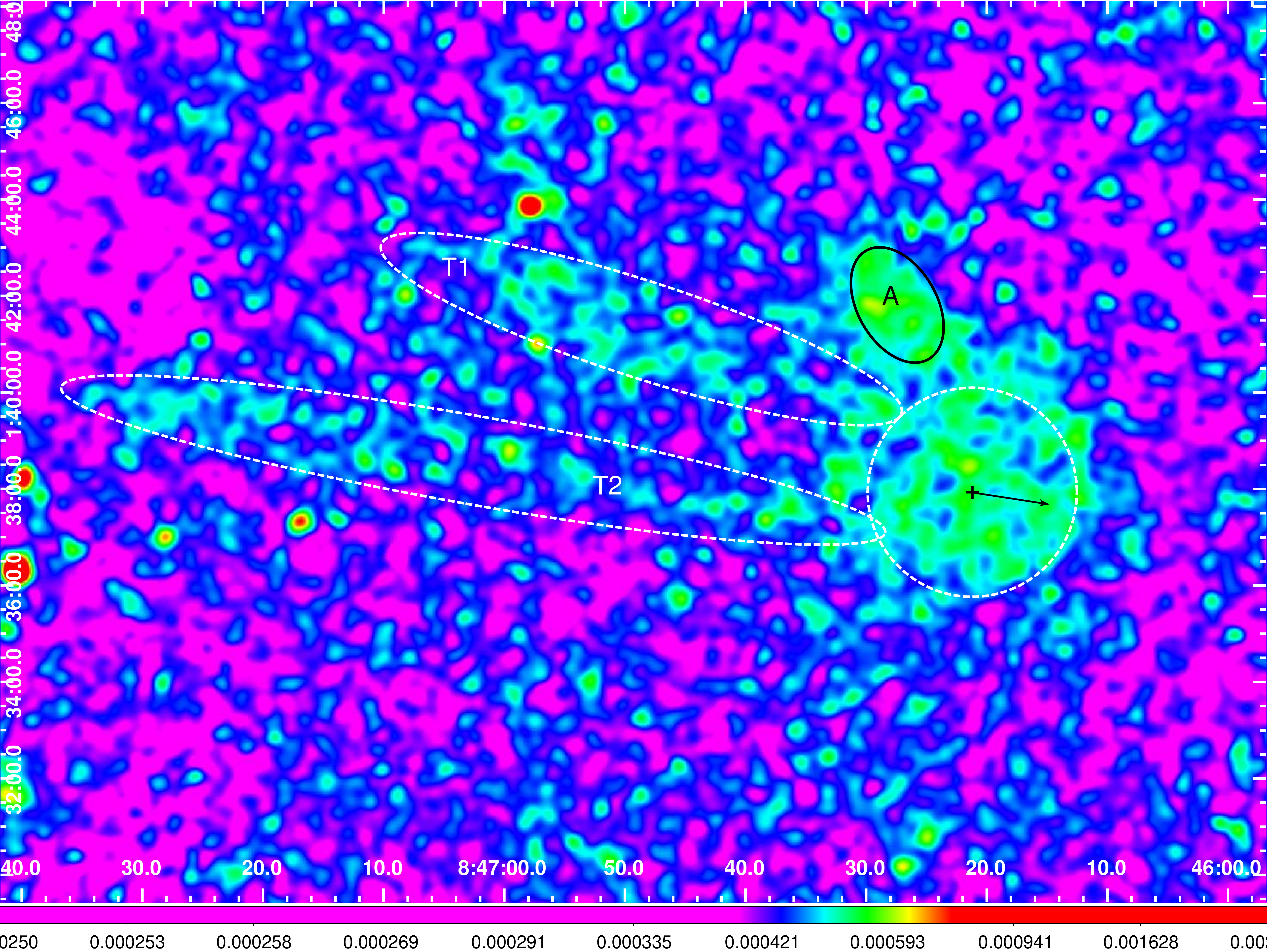}
\end{center}
\caption{The FUV emission towards EY\,Hya, imaged with GALEX. The large white dashed circle, with radius of 130${''}$, delineates the radial extent of the termination-shock in the astrosphere, west of EY\,Hya. Two long filamentary stuctures that comprise the tail region of the astrosphere are labelled T1 and T2. The emission in region labelled A is due to an imperfectly subtracted very bright point source located at an offset of $275{''}$ (at $PA=25^{\circ}$) from EY\,Hya. The black cross shows the star's location. Vector shows the star's proper motion of 9.5\,mas\,yr$^{-1}$ at $PA=261^{\circ}$, magnified by a factor 10$^4$. North is up and east is to the left. The scalebar shows intensity in cps/pix.}
\label{EYHyaFUV2}
\end{figure}

The FUV image of EY\,Hya (Fig.\,\ref{EYHyaFUV2}) shows a head-tail structure around the star. This structure has no counterpart in the NUV image. The long axis of this structure is closely aligned with EY\,Hya's proper motion vector\footnote{the proper motion for each source has been corrected for solar motion, see \S\,\ref{vstar}; the corrected values are used in all further reference to this parameter}. We therefore infer that the western periphery of the head represents the termination-shock of the astrosphere around EY\,Hya. The tail appears to consist of two long filamentary stuctures -- T1 (length$\sim340{''}$) and T2 (length$\sim520{''}$). The extended structure at $250{''}$ between $PA=25^{\circ}$ and $PA=40^{\circ}$ (Region A in Fig.\,\ref{EYHyaFUV2}) is not associated with EY\,Hya, but is due to an imperfectly subtracted very bright point source located at an offset of $275{''}$ (at $PA=25^{\circ}$) from EY\,Hya, that left some residual brightness after being removed. 

\begin{figure}[htbp]
\begin{center}
\includegraphics[width=16cm]{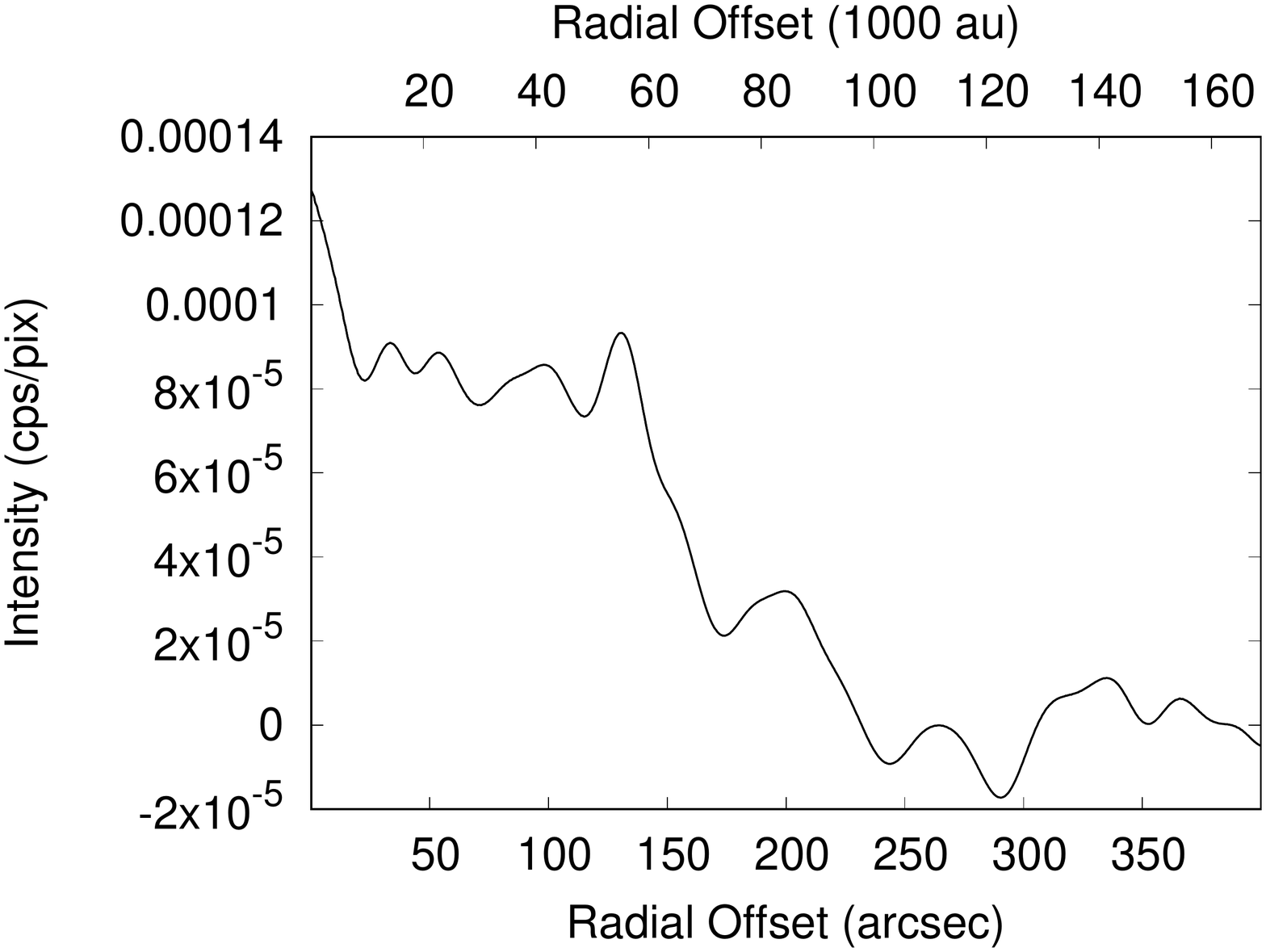}
\end{center}
\vspace{1.0in}
\caption{Radial intensity cut of the FUV emission around EY\,Hya, averaged over an azimuthal wedge with its apex centered on the star, and spanning the range from PA=-60\arcdeg~to -140\arcdeg. The average surrounding sky intensity has been subtracted from the cut.}
\label{eyhyacut}
\end{figure}

A radial intensity cut averaged over an azimuthal wedge with its apex centered on the star, and spanning the range from PA=-60\arcdeg~to -140\arcdeg~(Fig.\,\ref{eyhyacut}), shows a peak at radius $r=130{''}$, with a rapid decline of the intensity for larger radii out to $r\sim170{''}$. This is followed by a region of lower FUV intensity, out to $r\sim230{''}$ where its brightness becomes comparable to the surrounding sky intensity. We infer a termination-shock radius of $R_1=130{''}$, and an outer radius of the astropause, $R_c\sim230{''}$. The low-intensity region in the range $r\sim170{''}-230{''}$ likely represents emission from swept-up ISM material between the outer edge of the astropause and the bow-shock interface separating the shocked and unshocked ISM. 

\subsection{R\,LMi}
\begin{figure}[htbp]
\begin{center}
\includegraphics[width=16cm]{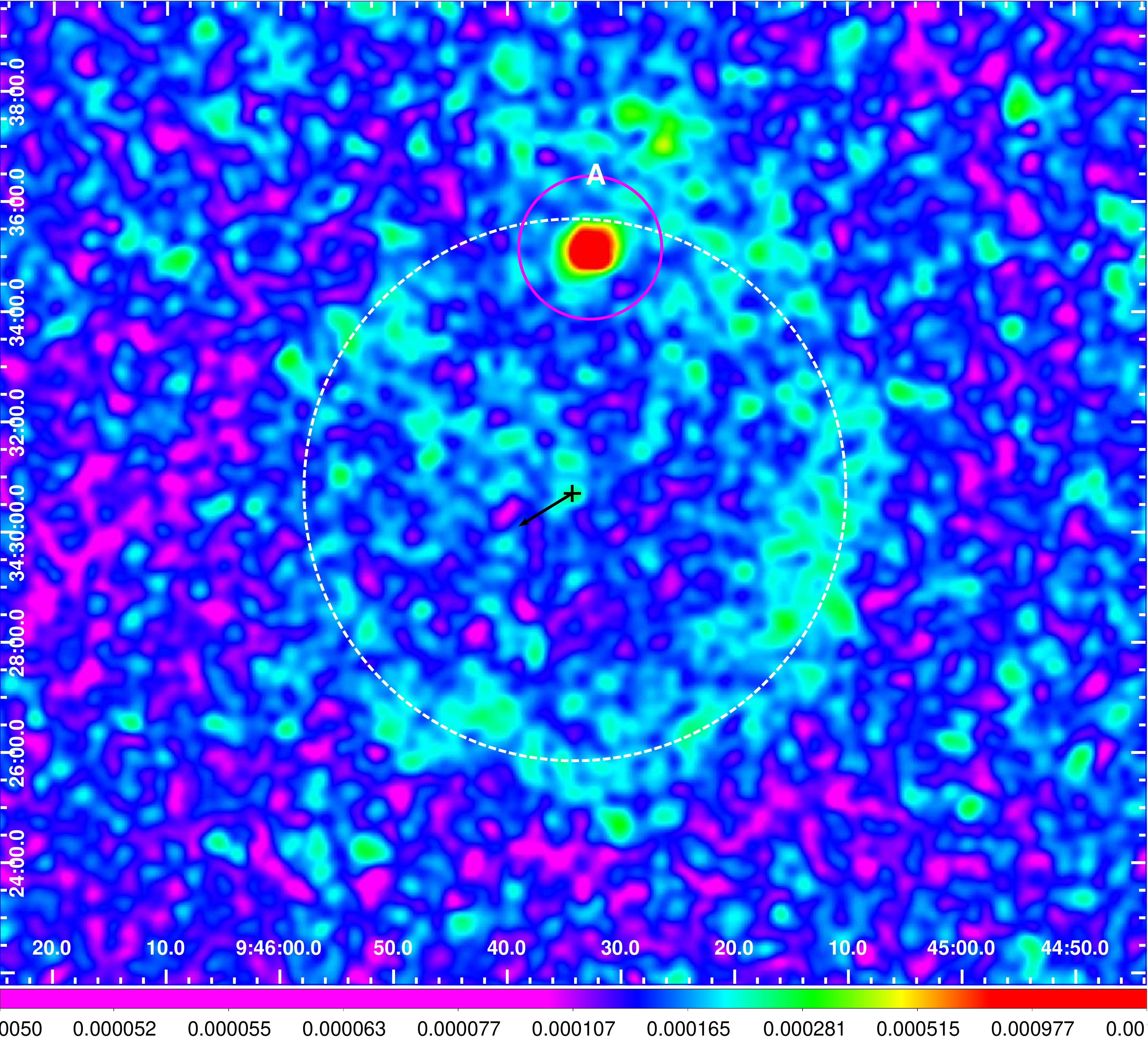}
\end{center}
\caption{The FUV emission towards R\,LMi, imaged with GALEX. The large white dashed circle, with radius of 295${''}$, delineates the radial extent of the termination-shock in the astrosphere; black cross shows location of the star. Vector shows the star's proper motion of 6.9\,mas\,yr$^{-1}$ at $PA=122^{\circ}$, magnified by a factor 10$^4$. Region A (magenta circle) shows local environment around two point sources that could not be removed. North is up and east is to the left.  The scalebar shows intensity in cps/pix.}
\label{RLMiFUV2}
\end{figure}

\begin{figure}[htbp]
\begin{center}
\includegraphics[width=16cm]{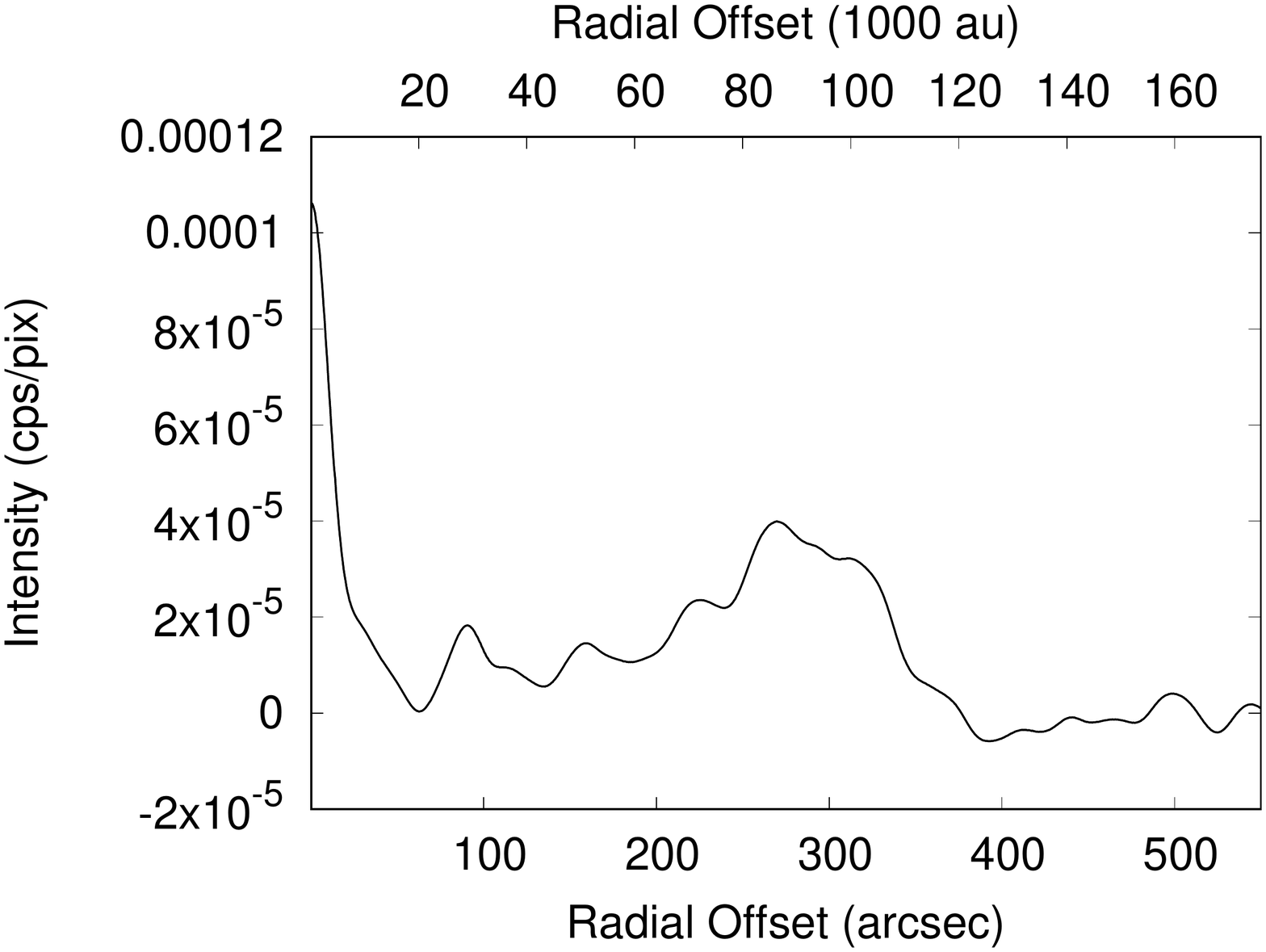}
\end{center}
\vspace{1.0in}
\caption{Radial intensity cut of the FUV emission around R\,LMi, averaged over an azimuthal wedge with its apex centered on the star, and spanning a 320\arcdeg~angular range around $PA=120\arcdeg$. The average surrounding sky intensity has been subtracted from the cut.}
\label{RLMiCuts}
\end{figure}

The FUV image of R\,LMi (Fig.\,\ref{RLMiFUV2}) shows a partial ring-like structure around the star. This extended FUV emission structure has no counterpart in the NUV image. A minor segment of the ring structure is overlapped by two bright point sources within the circle labeled A in the images that could not be removed.
The ring structure is somewhat brighter overall in a semicircular azimuthal wedge around $PA{\sim}-120\arcdeg$. A radial intensity cut averaged over an azimuthal wedge with its apex centered on the star, and spanning a 320\arcdeg~azimuthal range around $PA=120\arcdeg$ (Fig.\,\ref{RLMiCuts}), shows a broad, flat-topped peak at radius $r=295{''}$, with a full-width at half-maximum of $\sim90{''}$. The intensity declines rapidly beyond radius $r\gtrsim320{''}$, reaching the brightness level of the surrounding sky intensity at $r\gtrsim375{''}$. It is not clear exactly where the termination-shock resides within the broad intensity hump, as a compromise we assume that it resides in the middle, at $R_1=295{''}$; the outer radius of the astropause lies at $R_c\sim375{''}$. The proper motion of this star is relatively small (2.4\,mas\,yr$^{-1}$), consistent with the roughly circular shape of its astrosphere.

\subsection{U\,Ant}\label{uant-desc}
\begin{figure}[htbp]
\begin{center}
\includegraphics[width=16cm]{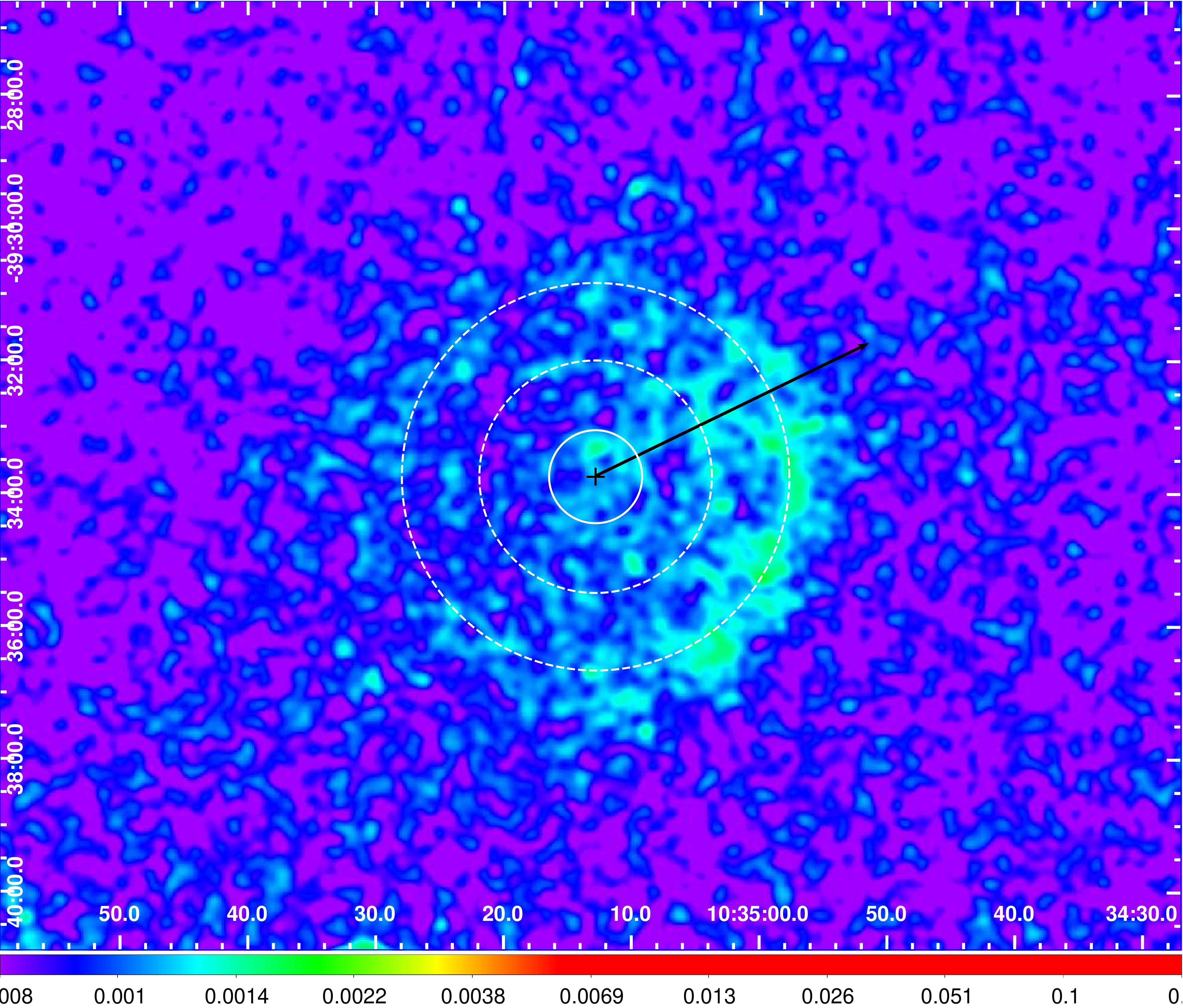}
\end{center}
\caption{The FUV emission towards  U\,Ant, imaged with GALEX. The large white dashed circle, with radius of 175${''}$, is a fit to the estimated radial location of the termination-shock in the astrosphere, seen to the west of U\,Ant. The smaller dashed circle, with radius of 105${''}$, marks the location of an intensity peak in the radial intensity cut in Fig.\ref{uantcut}. The small white dashed circle, with radius of (42${''}$) shows the location of the detached shell seen in the far-IR by \cite{Cox2012}. All circles are centered on the star's location (black cross). Vector shows the star's proper motion of $27.5$ mas\,yr$^{-1}$ at $PA=296^{\circ}$, magnified by a factor 10$^4$. North is up and east is to the left.  The scalebar shows intensity in cps/pix.}
\label{uantfuv}
\end{figure}

\begin{figure}[htbp]
\begin{center}
\includegraphics[width=16cm]{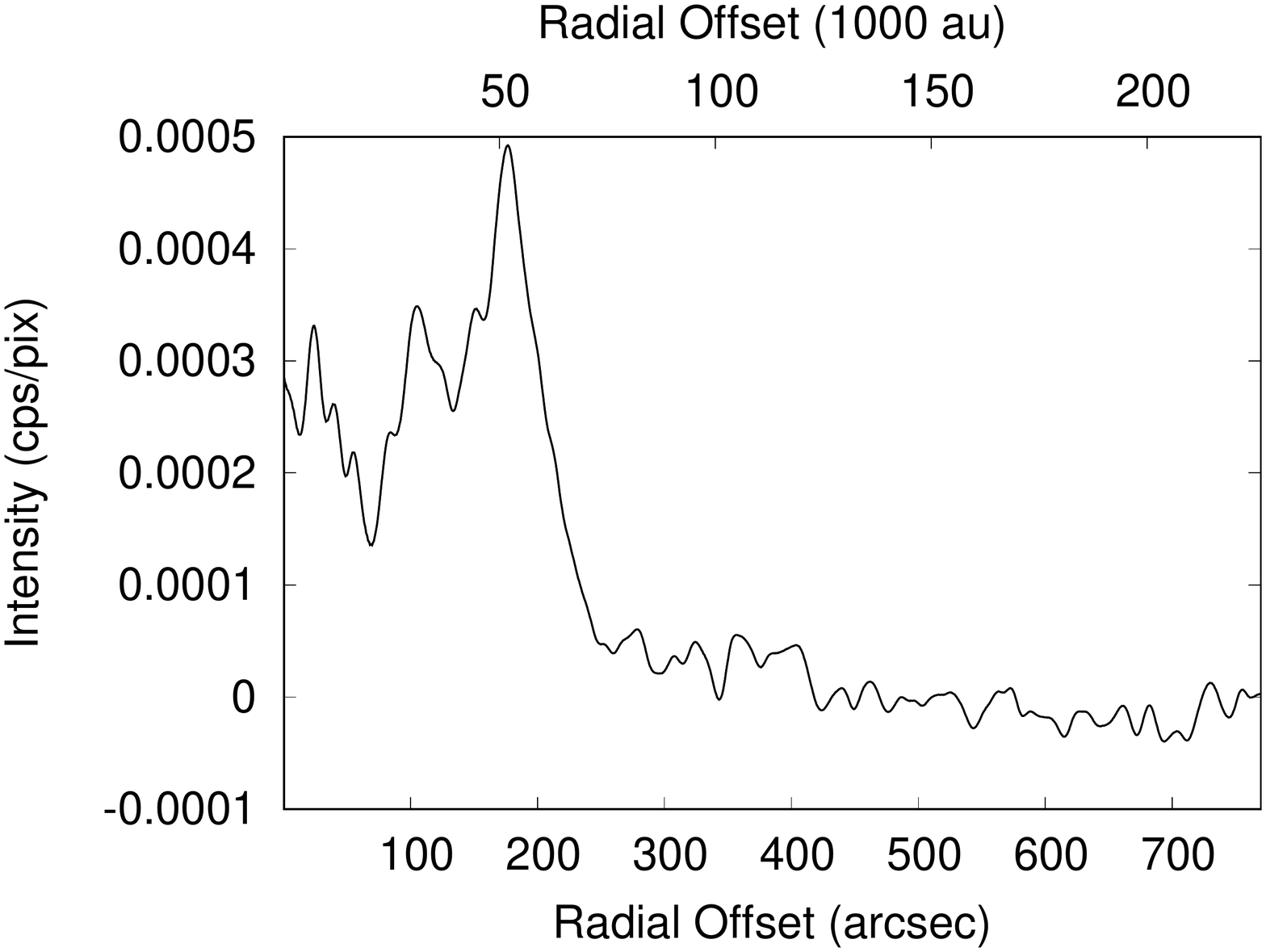}
\end{center}
\vspace{1.0in}
\caption{Radial intensity cut of the FUV emission around U\,Ant, averaged over an azimuthal wedge with its apex centered on the star, and spanning the range from PA=-60\arcdeg~to -140\arcdeg. The average  surrounding sky intensity has been subtracted from the cut.}
\label{uantcut}
\end{figure}

The FUV image of U\,Ant (Fig.\,\ref{uantfuv}) shows a ring-like structure around the star. This extended FUV emission structure has no counterpart in the NUV image. The FUV ring is much larger than the ring-like structure observed in the radial intensity cuts of Herschel/PACS images at a radius of $42{''}$ \citep{Kerschbaum2010,Cox2012}. Although the FUV emission can be seen all around the star, it is significantly brighter between position angles of $\sim-160\arcdeg$ to $\sim-40\arcdeg$, encompassing the direction of the star's proper motion. We infer that this bright region represents the termination shock of the astrosphere resulting from the interaction of star's wind with the ISM, and its symmetry axis is oriented at PA$\sim-100\arcdeg$. However, we note the possible presence of a very faint tail in the FUV emission that extends towards the south-west direction, suggesting that the symmetry axis of the astrosphere is at PA$\sim-45\arcdeg$. A radial intensity cut averaged over an azimuthal wedge with its apex centered on the star, spanning the range from PA=-60\arcdeg~to -140\arcdeg~(Fig. \ref{uantcut}), shows a prominent peak at radius $r=175{''}$, with a rapid decline of the intensity for larger radii out to $r\sim255{''}$. This is followed by a shallow decline in a region of low FUV intensity, out to $r\sim420{''}$, where its brightness becomes comparable to the surrounding sky intensity. We infer a termination-shock radius of $R_1=175{''}$, and an outer radius of the astropause, $R_c\sim255{''}$. The low-intensity region in the range $r\sim255{''}-420{''}$ likely represents emission from swept-up ISM material between the outer edge of the astropause and the bow-shock interface separating the shocked and unshocked ISM.

A secondary weaker peak is seen in FUV intensity cut at $r=105{''}$ -- the PACS 160\,\micron~imaging of U\,Ant shows faint patchy emission beyond $42{''}$ ring, but lack the sensivity to reveal a counterpart to the FUV peak. Since this peak lies interior to the termination-shock, we suggest that it is the result of a wind-wind interaction similar to the one that has been postulated for the $42{''}$ ring.

\cite{Izumiura1997} found, using IRAS 60\,\micron~and 100\,\micron~imaging, two extended dust shell structures in U\,Ant; they used these data to construct a double-shell model of this source. Their inner shell with a median radius of $51{''}$ likely corresponds to the shell seen in the PACS images; their outer shell, which is separated from the inner one by $150{''}$ likely corresponds to the astrosphere directly revealed in the FUV images.

\subsection{V\,Hya}\label{vhya-desc}
\begin{figure}[htbp]
\begin{center}
\includegraphics[width=16cm]{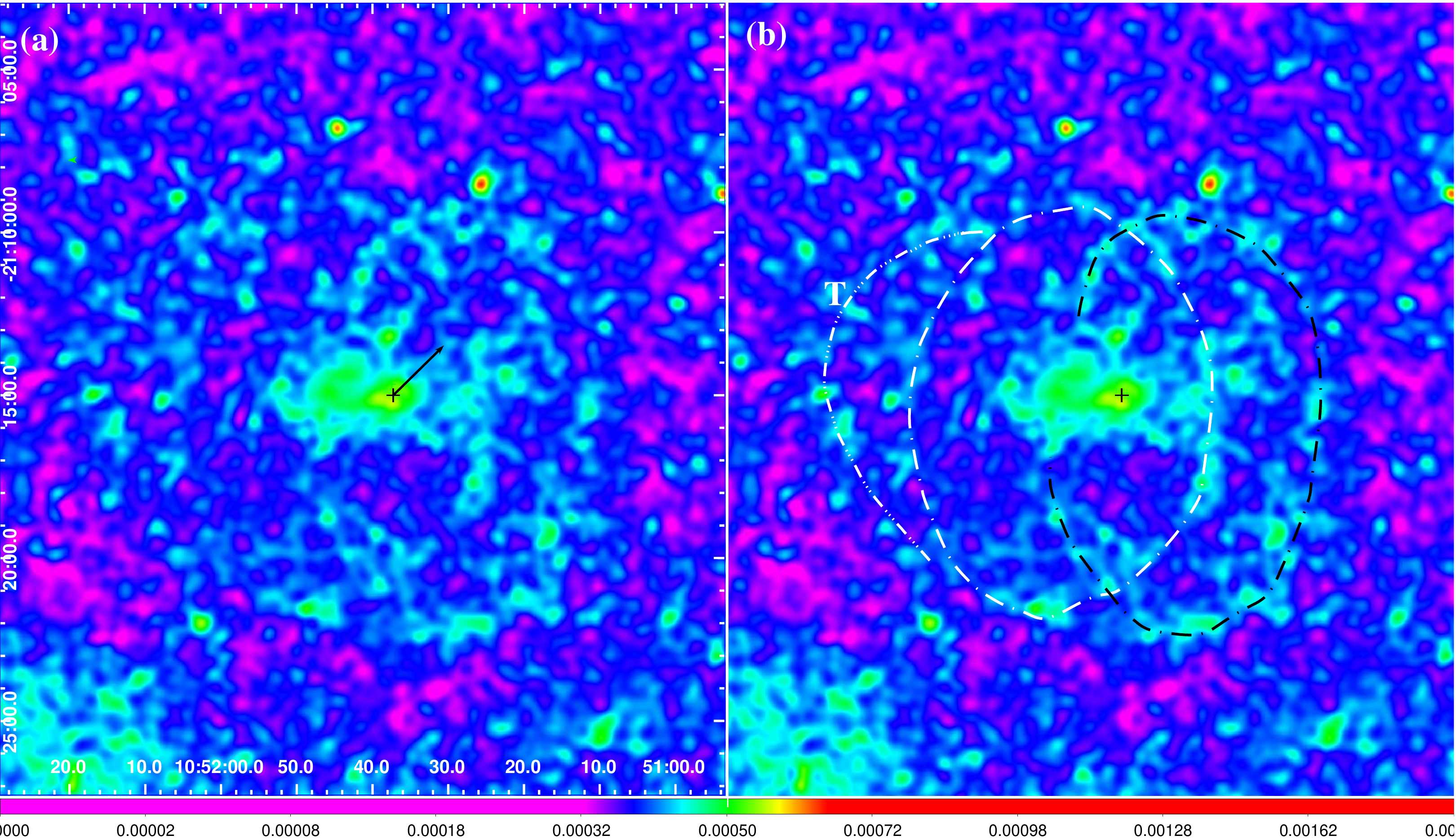}
\end{center}
\caption{The FUV emission towards V\,Hya, imaged with GALEX. In panel (a) black cross shows the location of the star, and vector shows the star's proper motion of $12.9$\,mas\,yr$^{-1}$ at $PA=315^{\circ}$, magnified by a factor 10$^4$. In panel (b), dashed-dotted  white) elliptical arcs (dashed black elliptical arc) delineate extended regions of FUV emission that are likely associated with the blue-shifted (red-shifted) large parabolic outflows seen in CO emission by \cite{Sahai2022}. Excess FUV emission (over the average sky background) is seen near the transverse structure labelled T. North is up and east is to the left.  The scalebar shows intensity in cps/pix.}
\label{VHyaFUV2}
\end{figure}

The FUV image of V\,Hya is complex, with a central elongated feature oriented east-west, and two large, roughly elliptical ring-like structures (Fig.\,\ref{VHyaFUV2}). These structures have no counterpart in the NUV image. The central elongated structure may be associated with the interacting of the extended extreme-velocity highly-collimated blobby outflows seen both in CO millimeter-line emission \citep{Sahai2022}) as well as optical line emission (\cite{Sahai2016, Scibelli2019}) with the ambient circumstellar medium.

The centers of the FUV emission rings lie along a roughly east-west axis. The ratio of major-to-minor axis of the rings is $\sim1.45$ -- assuming these are intrinsically circular, we conclude that the their axis is inclined by $\sim45\arcdeg$ to the sky-plane. A similar orientation has been found for the axes of the rings seen in the disk around V\,Hya via CO millimeter-line imaging (\cite{Sahai2022}). The FUV rings may be associated with the large high-velocity bipolar parabolic outflows seen in V\,Hya via CO millimeter-line imaging, whose axes are also aligned roughly east-west (\cite{Sahai2022}). In this scenario, the FUV emission would result from molecular H$_2$ interaction with the ISM or the circumstellar medium resulting from a spherical, slowly-expanding wind from V\,Hya. Some excess FUV emission (over the average sky background) is seen near the transverse structure labelled T.

The major axes of the FUV elliptical ring-structures is about $750{''}$, roughly a factor 35 larger than the widest extent of these outflows as seen in CO emission ($\sim20{''}$), showing that these outflows have been operating for a much longer time-scale ($t_{out}(FUV)$) than could be estimated from the CO data ($t_{out}(CO)$). A rough estimate of $t_{out}(CO)$ can be made from dividing the radial distance between the star and the tip of the outflow seen in Fig.\,13d of \cite{Sahai2022} ($16{''}\times311$\,pc/cos\,$i$=7035\,au), where $i\sim45\arcdeg$ is the inclination angle of the outflow axis to the sky-plane, by an estimate of the expansion velocity of the material there, $V_{out}$, assuming radial expansion. Correcting for projection, the radial velocity offset of the CO emission seen in Fig.\,13d of \cite{Sahai2022} at $47$\,\kms~from the systemic velocity ($-17.4$\,\kms) of 64.4\,\kms~implies a 3D expansion velocity of $\sim185$\,\kms, which results in an expansion age of $t_{out}(CO)\sim180$\,yr. Assuming the same expansion velocity for the material seen in the FUV elliptical ring, but located at a projected radial distance that is about a factor 25 larger than that seen in CO, we find an expansion age of $t_{out}(FUV)\gtrsim4500$\,yr. This age is a lower limit because the material seen in FUV emission must have slowed down due its interaction with the ambient medium.

Because the FUV rings have elliptical shapes, the use of a large angular-wedge for a radial intensity cut signficantly dilutes the intensity peak produced by the rings. Hence we have made a $20\arcdeg$-wide cut around $PA=-90\arcdeg$ to estimate the radius of the western outflow's interaction with the ISM. The cut (Fig.\,\ref{VHyaFUV2}) shows a peak at $r\sim365{''}$ that corresponds to the western-most part of the elliptical ring centered the west of the star (dashed-black ring in Fig.\,\ref{VHyaFUV2}), along its minor axis. A less prominent hump is seen at $r\sim135{''}$ that corresponds to the western-most part of the elliptical ring centered east of the star (dashed-dotted white ring in Fig.\,\ref{VHyaFUV2}).
\label{VHyaCuts}

There is no obvious indication of the interaction of a spherical wind with the ISM. 

\subsection{RT\,Vir}
\begin{figure}[htbp]
\begin{center}
\includegraphics[width=16cm]{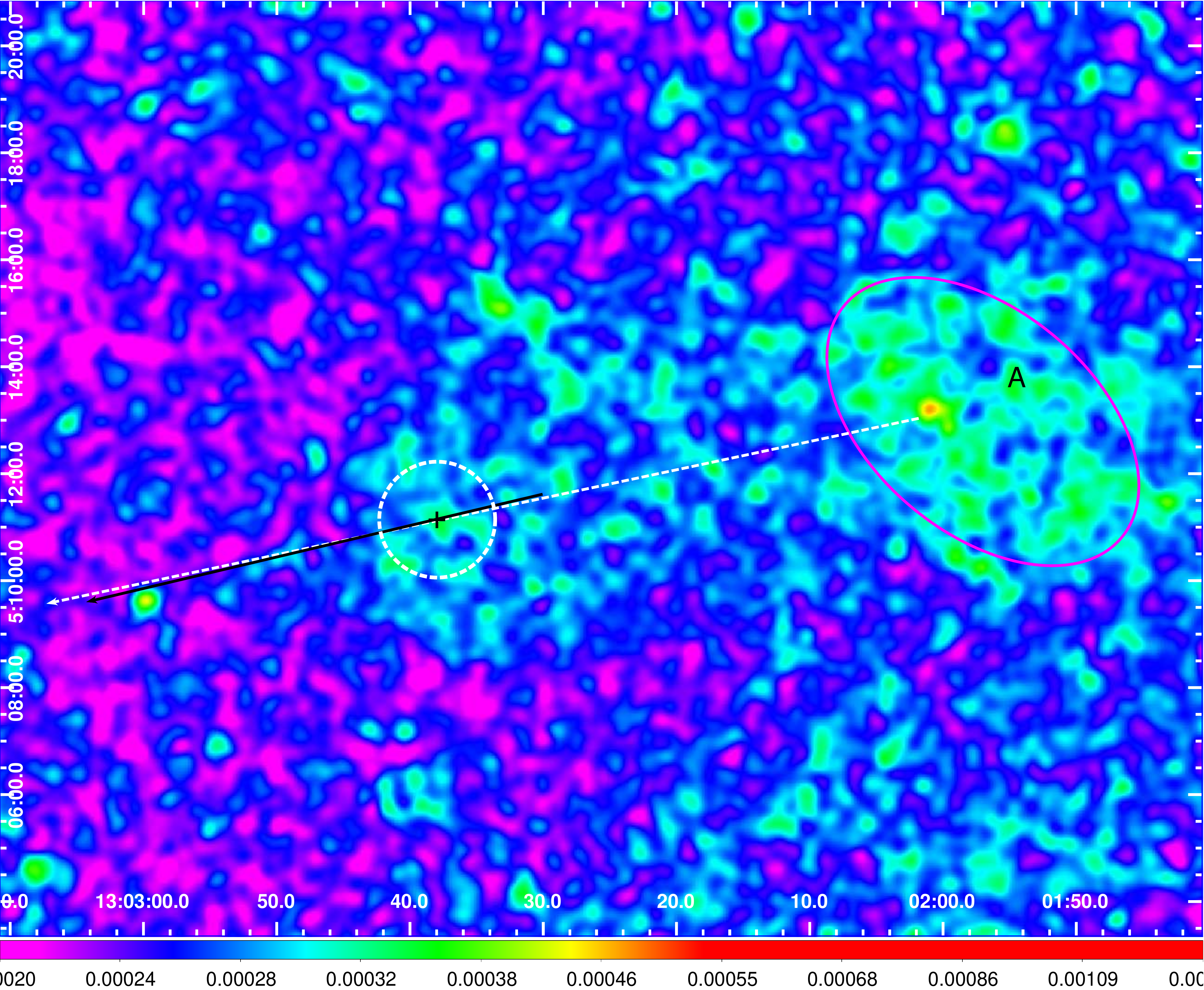}
\end{center}
\caption{The FUV emission towards RT\,Vir, imaged with GALEX. The white dashed circle, with radius of 65${''}$, delineates the termination-shock in the astrosphere, east of the star (black cross). Vector shows the star's proper motion of $52.6$\,mas\,yr$^{-1}$ at $PA=103^{\circ}$, magnified by a factor 10$^4$. Region A (magenta ellipse) shows a locally bright region near the end of the astrosphere tail. North is up and east is to the left.  The scalebar shows intensity in cps/pix.}
\label{RTVirFUV2}
\end{figure}

\begin{figure}[htbp]
\begin{center}
\includegraphics[width=16cm]{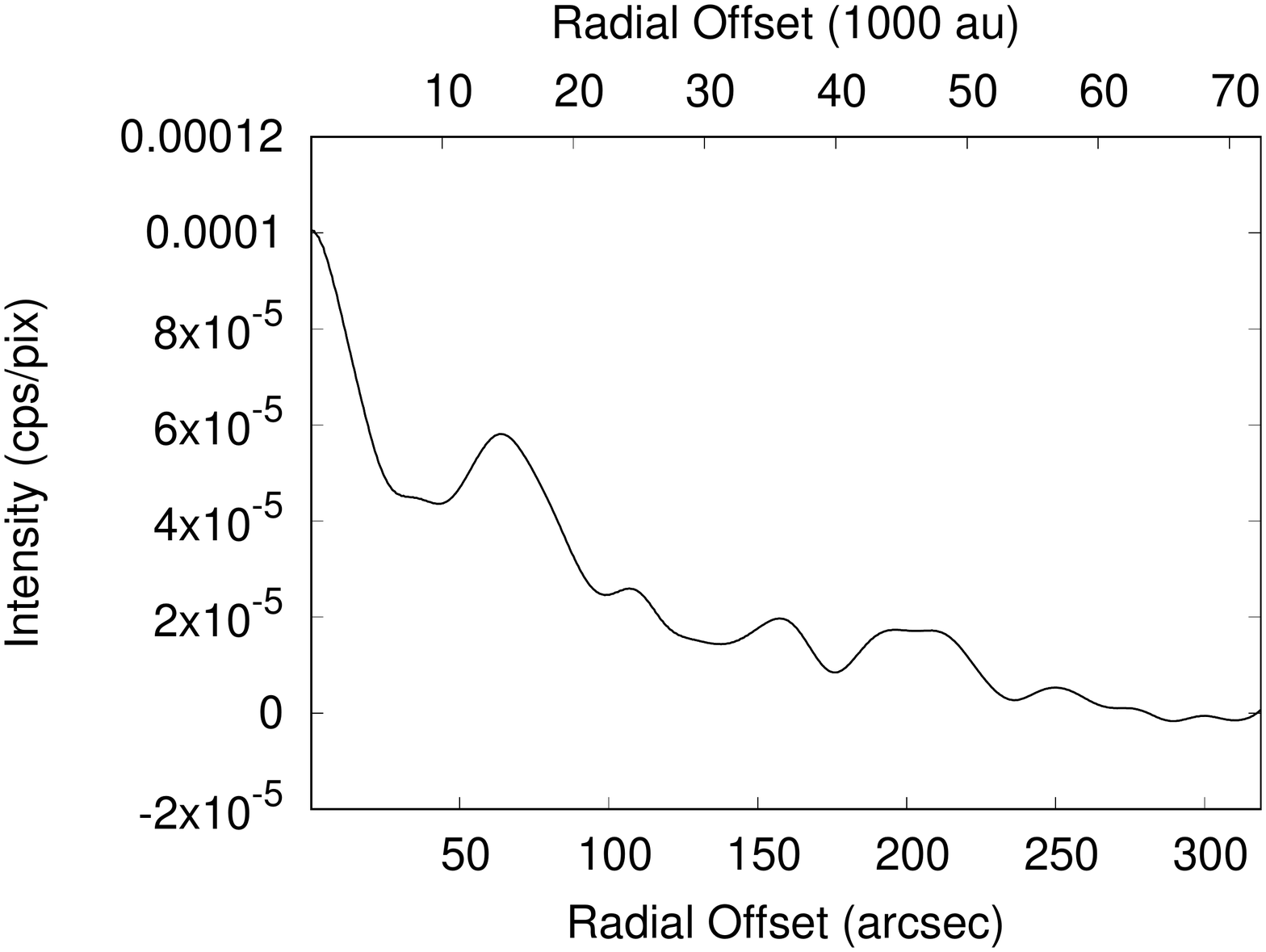}
\end{center}
\vspace{1.0in}
\caption{Radial intensity cut of the FUV emission around RT\,Vir, averaged over an azimuthal wedge with its apex centered on the star, and spanning the range from PA=65\arcdeg~to 145\arcdeg. The average surrounding sky intensity has been subtracted from the cut.}
\label{RTVirCuts}
\end{figure}

The FUV image of RT\,Vir shows an elongated head-tail structure, with a bow-shock like shape east of the star, and an extended tail to its west. This extended FUV emission structure has no counterpart in the NUV image. The long axis of this structure is aligned along $PA\sim192\arcdeg$, i.e., very close to that of the proper-motion vector (Fig.\,\ref{RTVirFUV2}). We therefore infer that this structure represents the termination-shock of the astrosphere around RT\,Vir. A radial intensity cut averaged over an azimuthal wedge with its apex centered on the star, and spanning the range from PA=65\arcdeg~to 145\arcdeg~(Fig.\,\ref{RTVirCuts}), shows a peak at radius $r=65{''}$, with a fast decline of the intensity for larger radii out to $r\sim95{''}$. This is followed by a shallower decline in a region of low FUV intensity, out to $r\gtrsim240{''}$, where its brightness becomes comparable to the surrounding sky intensity. We infer a termination-shock radius of $R_1=65{''}$; and an outer radius of the astropause, $R_c\sim95{''}$. The low-intensity region in the range $r\sim95{''}-240{''}$ likely represents emission from swept-up ISM material between the outer edge of the astropause and the bow-shock interface separating the shocked and unshocked ISM.

There is a locally bright extended structure near the end of the tail (Region A) that may or may not belong to the astrosphere. 

\subsection{R\,Hya}
\begin{figure}[htbp]
\begin{center}
\includegraphics[width=16cm]{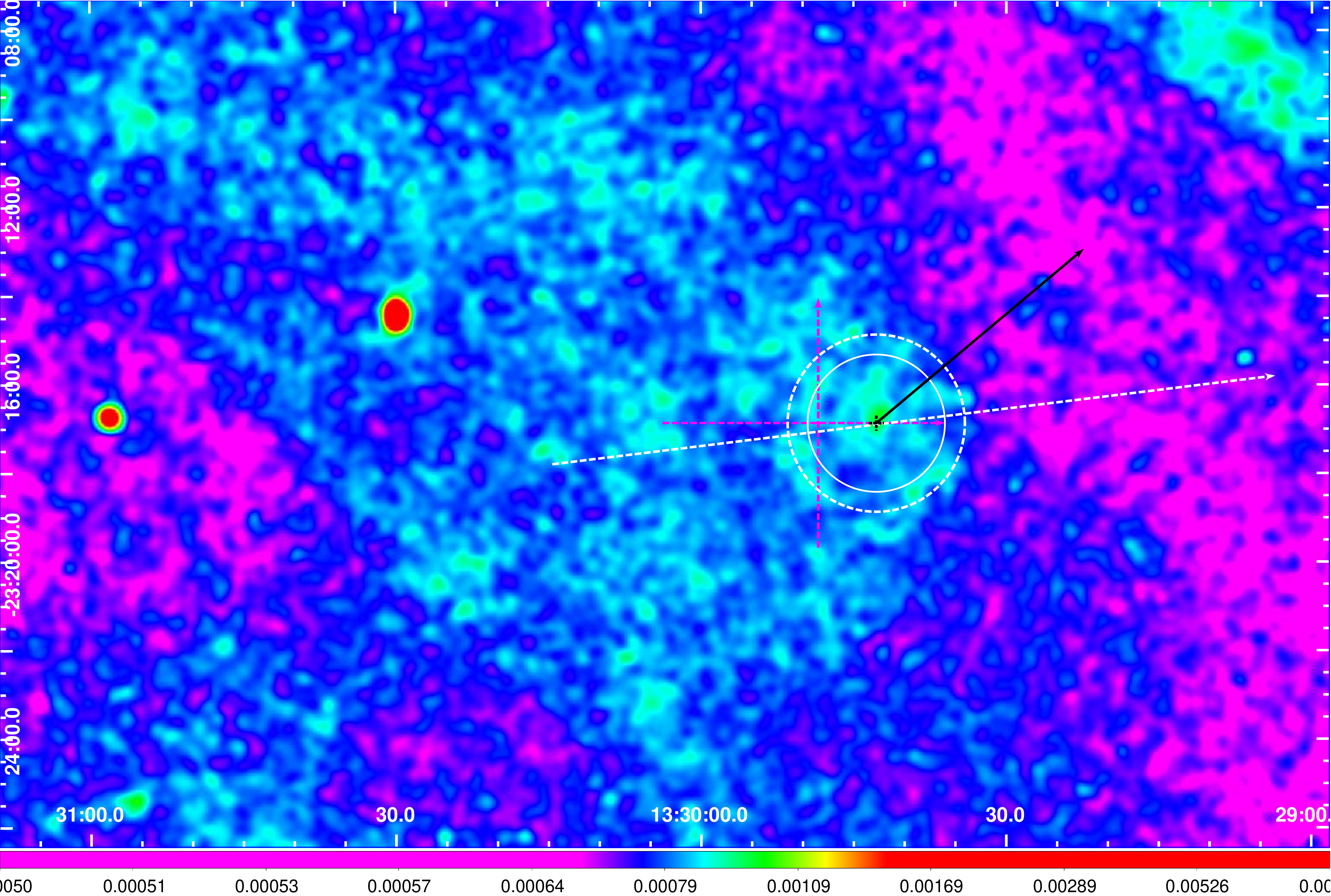}
\end{center}
\caption{The FUV emission towards R\,Hya, imaged with GALEX. The white dashed circle, with radius of 120${''}$, delineates the radial extent of the termination-shock in the astrosphere, west of R\,Hya (black cross); white solid circle delineates a termination-shock radius of 93${''}$ estimated from far-IR imaging by \cite{Cox2012}. Black vector shows the star's proper motion of $36.7$\,mas\,yr$^{-1}$ at $PA=310^{\circ}$, magnified by a factor 10$^4$; white vector shows symmetry axis of the fan-shaped astrosphere. North is up and east is to the left.  The scalebar shows intensity in cps/pix.}
\label{RHyaFUV2}
\end{figure}

The FUV image of R\,Hya shows an elongated fan-shaped structure, with a bow-shock like shape to the west of the star (Fig.\,\ref{RHyaFUV2}). This extended FUV emission structure has no counterpart in the NUV image. The symmetry axis of this structure is aligned along $PA\sim-82\arcdeg$, i.e., within $32\arcdeg$ of the proper-motion vector; we infer that this structure represents the astrosphere around this star. A similar wide fan-shaped structure has been seen in the far-infrared at 70\,\micron~with Spitzer (\cite{Ueta2006}) and 70 and 160\,\micron~withPACS (\cite{Cox2012}), but appears to be much more limited in extent compared to the FUV fan structure. The maximum north-south (east-west) extent of the fan structure as seen in the PACS 70\,\micron~image is $\sim340{''}$ ($\sim385{''}$), as denoted by dashed vertical (horizontal) magenta vectors in Fig.\,\ref{RHyaFUV2} -- these vectors lie well within the FUV emission structure. \cite{Cox2012} derive a termination-shock radius of $93{''}$ from their far-IR imaging.

\begin{figure}[htbp]
\begin{center}
\includegraphics[width=16cm]{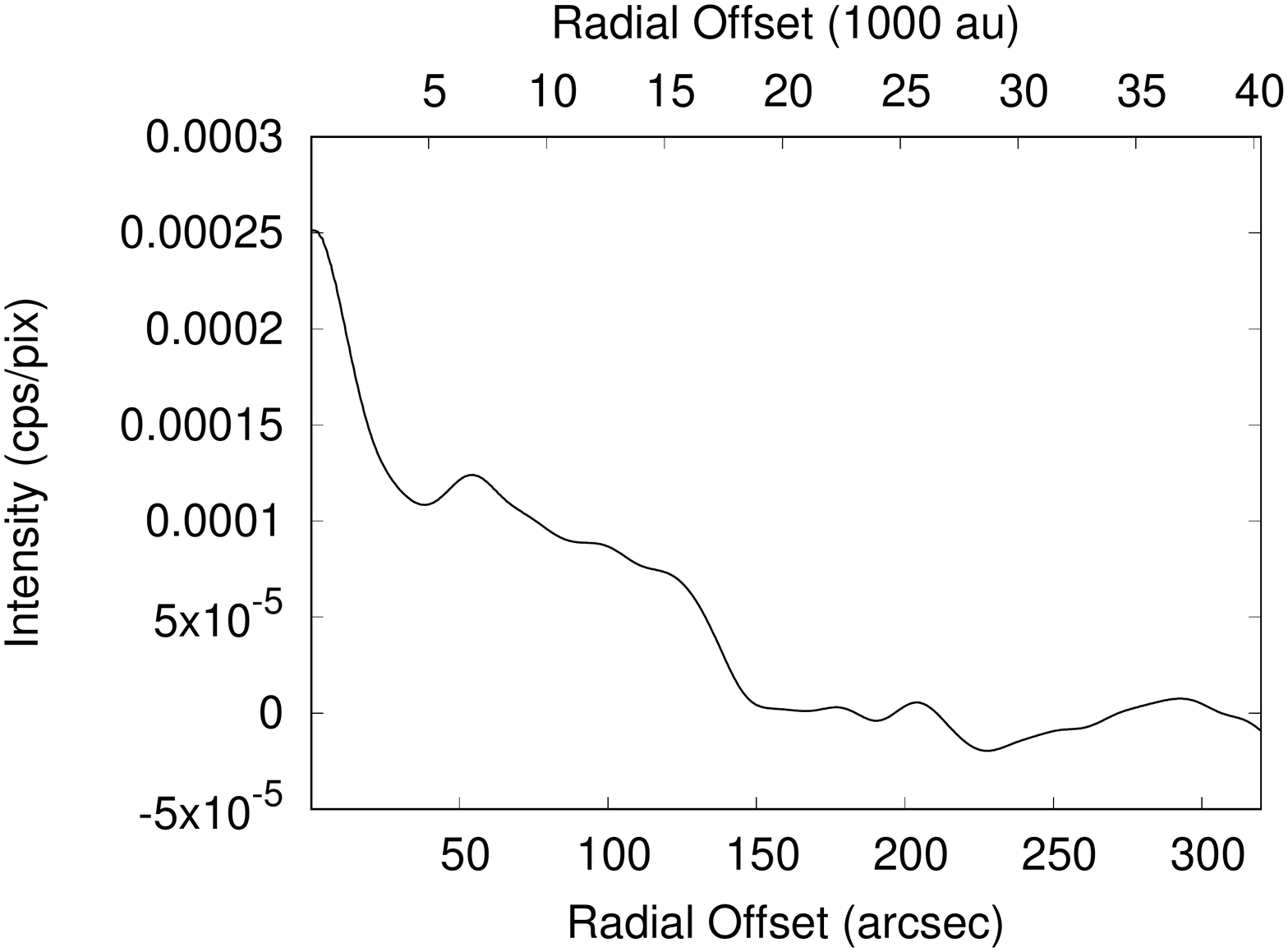}
\end{center}
\vspace{1.0in}
\caption{Radial intensity cut of the FUV emission around R\,Hya, averaged over an azimuthal wedge with its apex centered on the star, and spanning the range from $PA=-123$\arcdeg~to $-43$\arcdeg. The average surrounding sky intensity has been subtracted from the cut.}
\label{RHyaCuts}
\end{figure}

A radial intensity cut of the FUV emission around R\,Hya, averaged over an azimuthal wedge (spanning the range from $PA=-123$\arcdeg~to $-43$\arcdeg) with its apex centered on the star (Fig.\,\ref{RHyaCuts}), show a steep decline in the intensity starting at $r\gtrsim120{''}$ and reaching the background sky level at $r\sim145{''}$. We infer that the termination-shock radius is $R_1=120{''}$, which is larger than the value derived from the far-IR data. Narrowing the opening angle of the azimuthal wedge to $40\arcdeg$, makes no significant difference to the radial intensity distribution. We do not find any indication of a dramatic change in the radial FUV intensity at $\sim93{''}$. We infer an outer radius of the astropause, $R_c\sim145{''}$. 

\subsection{W\,Hya}
\begin{figure}[htbp]
\begin{center}
\includegraphics[width=16cm]{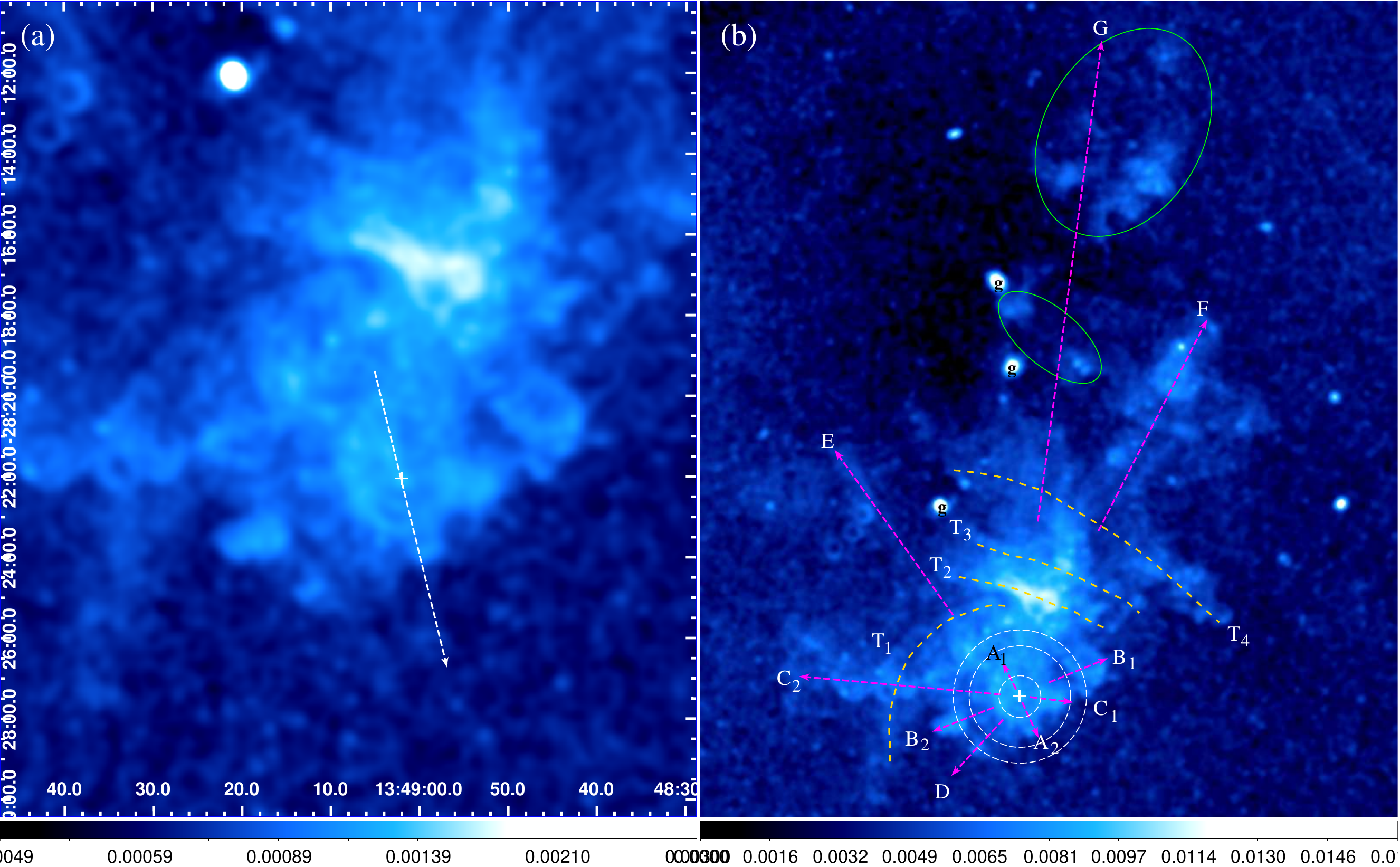}
\end{center}
\caption{The FUV emission towards W\,Hya (white cross), imaged with GALEX. (a) The FUV emission in the near vicinity of W\,Hya (white cross); white dashed vector shows the star's proper motion of $45.4$\,mas\,yr$^{-1}$ at $PA=194^{\circ}$, magnified by a factor 10$^4$. (b) The FUV emission around W\,Hya over a large field-of-view, with major structural features marked. Dashed white circles show partial ring-like structures of radii $65{''}$, $160{''}$, and $220{''}$. Several radial (compact as well as extended) features, likely corresponding to outflows are labelled as A, B, ... G. Outflows A, B, and C show diametrically-opposed pairs (denoted with subscripts 1 and 2, e.g., A$_1$ and  A$_2$). Outflow G is inferred from the presence of diffuse emission regions (encircled with green ellipses) located roughly north of the star. Four large azimuthal structures are labelled T$_1$--T$_4$. Bright compact bright sources (labelled ``g") are likely background galaxies.   North is up and east is to the left.  The scalebar shows intensity in cps/pix.}
\label{WHyaFUV2}
\end{figure}

The FUV image of W\,Hya shows an overall head-tail morphology, but with an amazingly extensive and complex structure (Fig.\,\ref{WHyaFUV2}), that includes both azimuthal and radial features. The NUV image of W\,Hya does not show these structures. Amongst the azimuthal features, there are partial ring-like structures that produce noticeable intensity peaks in a radial intensity cut, averaged over an azimuthal wedge with its apex centered on the star, covering the azimuthal range $PA=95\arcdeg$ to $235\arcdeg$ (this range is selected to be as wide as possible but still avoid bright radial features; Fig.\,\ref{WHyaFUV2}). A strong, sharp intensity peak at $r=65{''}$ is due to the presence of a ring of the same radius, also seen in a 70\,\micron~PACS images (\cite{Cox2012}). This is followed by a weaker, but sharp intensity peak at $r=160{''}$, and a broader, asymmetric peak with its centroid at $r\sim220{''}$. There are counterparts to each of these peaks in the 160\,\micron~PACS radial intensity cuts; \cite{Cox2012} list only the outer one (at $r\sim230{''}$) in their study. These intensity peaks correspond to partial ring-like structures in the FUV image. There are several large azimuthal structures at larger radial distances (marked T$_1$--T$_3$).

Numerous radial features (both close to and far away from star) are also seen in the FUV image -- some of these may represent collimated outflows. These are labelled as A, B, ... G. Features A, B, and C show diametrically-opposed pairs (denoted with subscripts 1 and 2, e.g., A$_1$ and  A$_2$), i.e. and may be bipolar outflows. The tail region extends towards the north-north-west (features F and G). We find radial features in the PACS 70 and 160\,\micron~images of W\,Hya from the \cite{Cox2012} study, located at position angle similar to those of outflows A$_1$, A$_2$, (B$_2$+D), (C$_1$+B$_1$) and (F+G)\footnote{we have grouped together the FUV outflows that appear as one azimuthally-broad feature in the far-IR. Features F\,+\,G cannot be distinguished from the bright FUV emission in the near-vicinity of the star, but we find an azimuthally-broad far-IR feature in this region that roughly spans the azimuthal range covered by F and G}. \cite{Cox2012} mention only the counterpart to the A$_1$ outflow (as a jet-like extension pointing roughly north-east), but not the other radial features.

We infer that the outermost (rather broad) radial intensity peak at $r\sim220{''}$ corresponds to the interaction of W\,Hya's wind with the ISM. A plausible explanation of the innermost peak at $r=65{''}$ is that it results from the interaction of a higher-expansion velocity episode of enhanced mass-loss interacting with an older, lower-density, slower wind. Such a modification of mass-loss rate and expansion velocity can result from a thermal pulse, and is the preferred explanation for the presence of the ``detached" shells seen in a number of carbon stars (\cite{Olofsson2010} and references therein). The ring at $r=160{''}$ may have a similar origin as the one at $r=65{''}$.

\subsection{RW\,Boo}
\begin{figure}[htbp]
\begin{center}
\includegraphics[width=16cm]{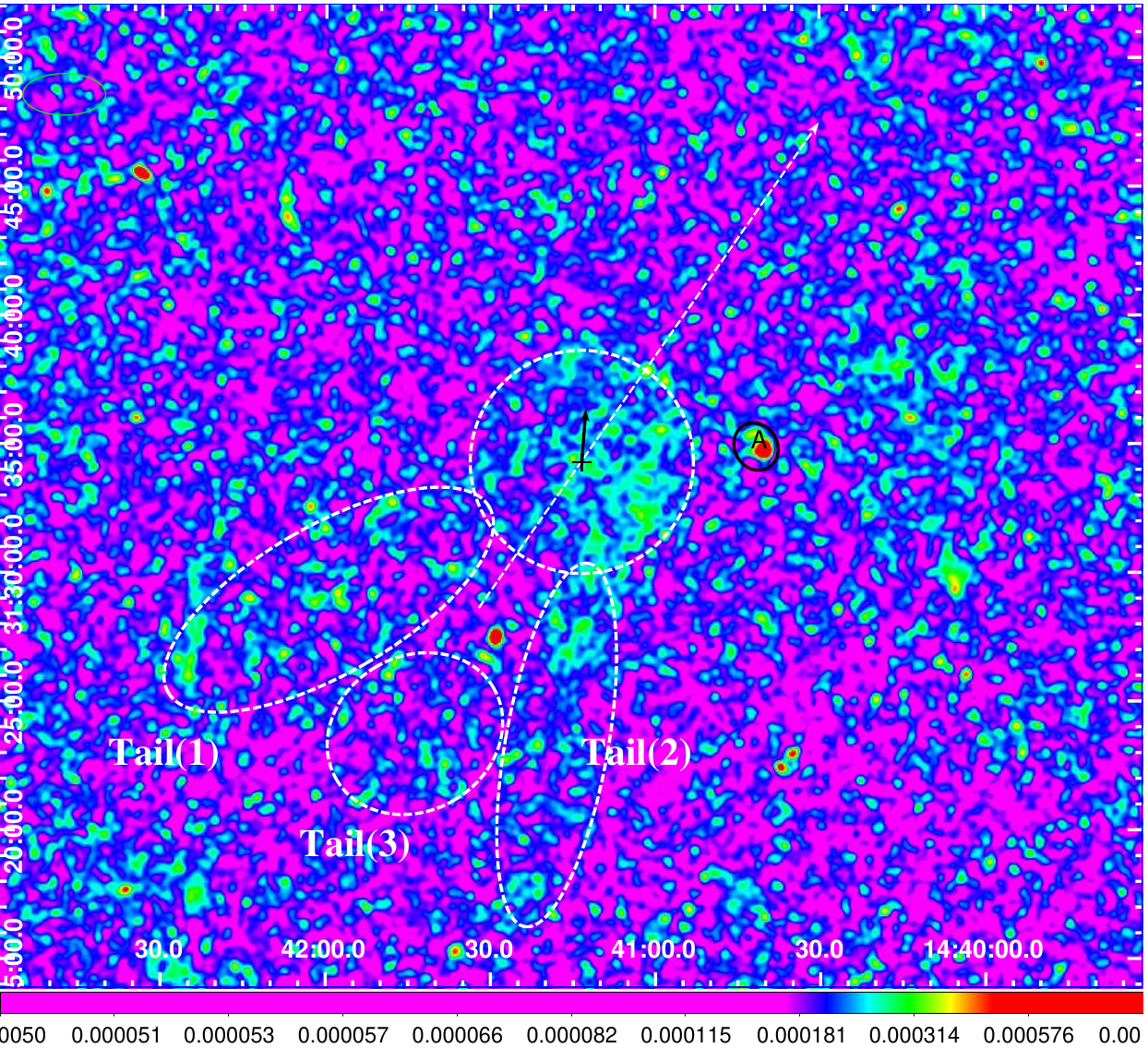}
\end{center}
\caption{The FUV emission towards RW\,Boo, imaged with GALEX. The dashed white circle (radius $260{''}$), delineates the radial extent of the termination-shock in the astrosphere, seen to the west of RW\,Boo. Black vector shows the star's proper motion of $12.4$\,mas\,yr$^{-1}$ at $PA=356^{\circ}$, magnified by a factor 10$^4$; white vector shows symmetry axis of the astrosphere.  Region A (black ellipse) shows local environment around a bright point source that could not be adequately removed.  North is up and east is to the left.  The scalebar shows intensity in cps/pix.}
\label{RWBooFUV2}
\end{figure}

The FUV image of RW\,Boo (Fig.\,\ref{RWBooFUV2}) shows relatively bright emission in a $\sim$180\arcdeg~wedge around $PA\sim-90\arcdeg$, suggesting that the termination-shock lies west of RW\,Boo with a east-west symmetry axis. This extended FUV emission structure has no counterpart in the NUV image. However, we also see faint tail structures south-east of the star, characterised by enhanced emission in patches labeled Tail(1), Tail(2) and Tail(3), suggesting that the symmetry axis of the astrosphere lies along $PA\sim-35\arcdeg$. However, the proper motion vector is directed along $PA=-4\arcdeg$, closer to the orientation of Tail(1), suggesting that the latter defines orientation of the symmetry axis.

Assuming that the symmetry axis is one that bisects the tail region, we have made a radial intensity cut averaged over an azimuthal wedge with is apex centered on the star, spanning the range from $PA=-75$\arcdeg~to $5$\arcdeg~(Fig.\,\ref{RWBooFUV2}).
The cut shows a sharp peak at $r=260{''}$ followed by a steep decline to the average sky background intensity at $r\sim310{''}$ -- we therefore infer a termination shock radius, i.e., $R_1=260{''}$, and an astrosheath outer radius of $R_c=310{''}$. Radial intensity cuts assuming an east-west symmetry axis yield similar results.

There is a possible low-intensity region in the range $r\sim310{''}-425{''}$ that likely represents emission from swept-up ISM material between the outer edge of the astropause and the bow-shock interface separating the shocked and unshocked ISM.

\subsection{RX\,Boo}
\begin{figure}[htbp]
\begin{center}
\includegraphics[width=16cm]{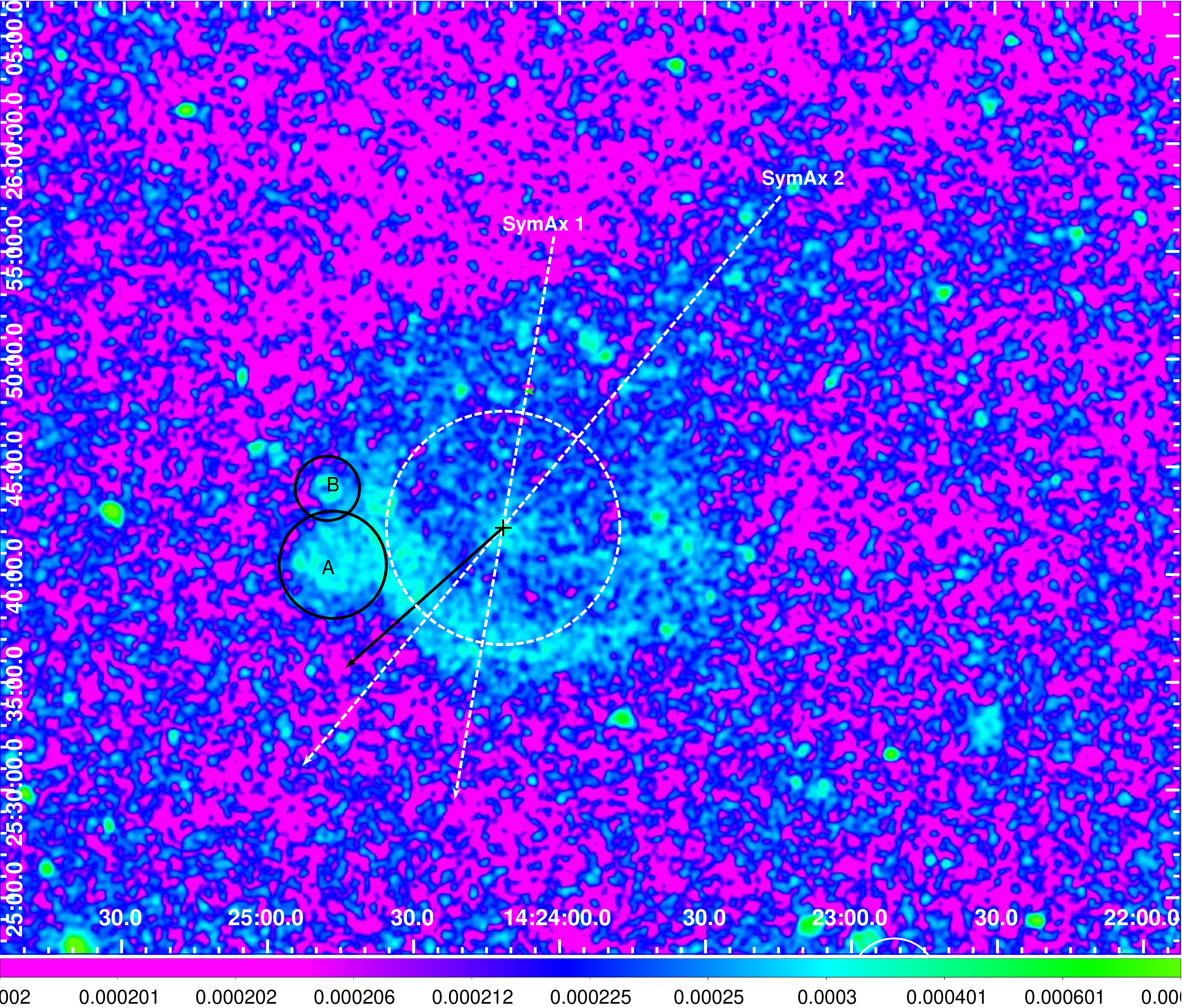}
\end{center}
\caption{The FUV emission towards RX\,Boo, imaged with GALEX. The white dashed circle (radius 325${''}$), delineates the radial extent of the termination-shock in the astrosphere, south-south-east of RX\,Boo (black cross). The white dashed vectors labelled SymAx 1 and SymAx2 represent two possible symmetry axes of the astrosphere. Black vector shows the star's proper motion of $58.8$\,mas\,yr$^{-1}$ at $PA=132^{\circ}$, magnified by a factor 10$^4$. A shows where a bright point source could not be removed; B shows a related ghost artifact. North is up and east is to the left.  The scalebar shows intensity in cps/pix.}
\label{RXBooFUV2}
\end{figure}

The FUV image of RX\,Boo (Fig.\,\ref{RXBooFUV2}) shows a bow shock structure to the south-south-east side of the star. The symmetry axis of this structure is somewhat uncertain -- one possibility is SymAx\,1, at PA$\sim170\arcdeg$ which bisects the ``nose" structure in the bow-shock; a second possibility is SymAx\,2, at PA$\sim140\arcdeg$ which bisects a spiny structure north-west of the star that may be the astrosphere's tail. The proper motion vector lies much closer to SymAx2 than SymAx1, suggesting that SymAx1 is the symmetry axis of the astrosphere.

\begin{figure}[htbp]
\begin{center}
\includegraphics[width=16cm]{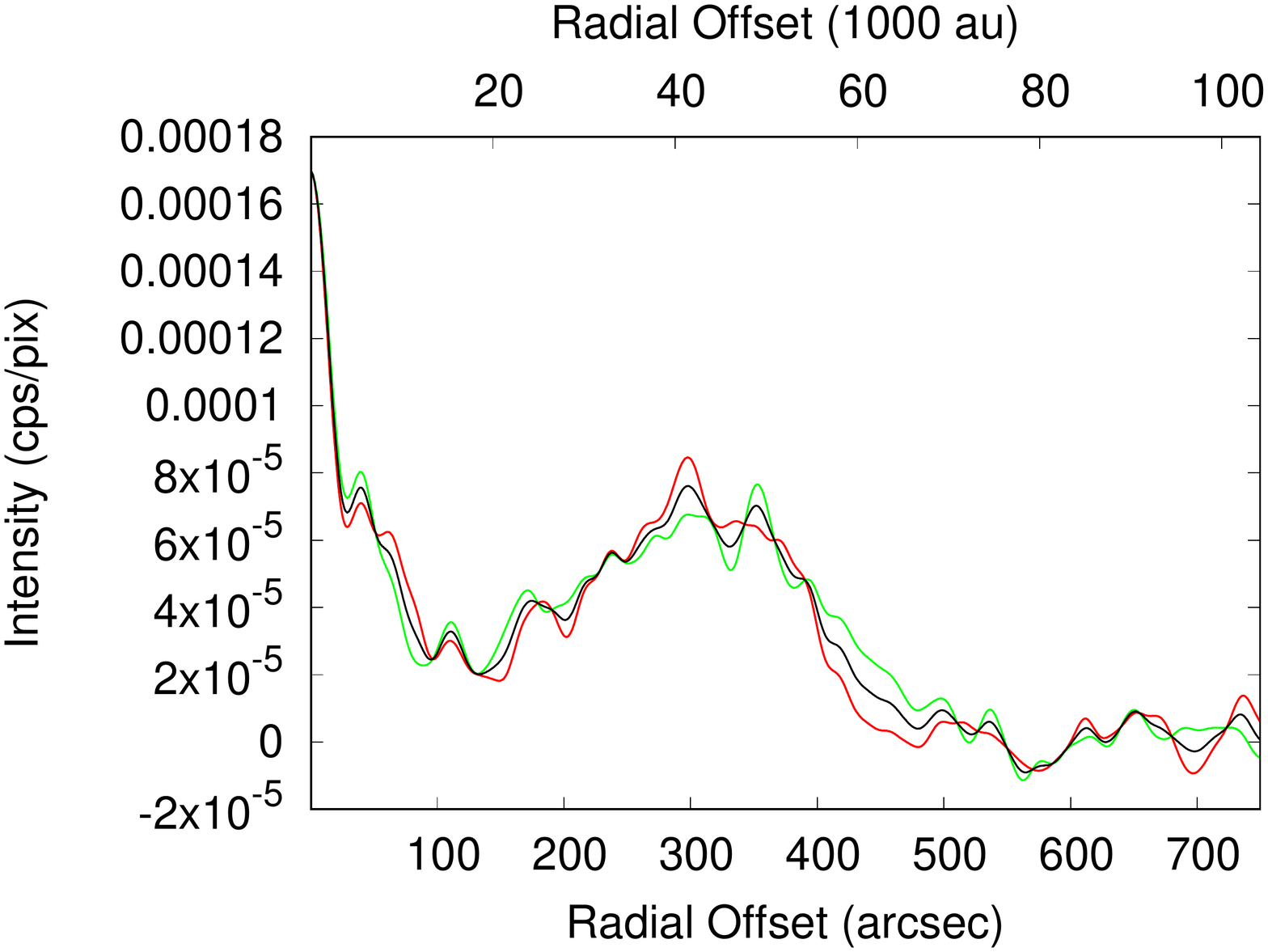}
\end{center}
\vspace{1.0in}
\caption{Radial intensity cuts of the FUV emission around RX\,Boo, averaged over azimuthal wedges with their apexes centered on the star, and spanning the range from $PA=150$\arcdeg~to $190$\arcdeg~(green curve), $PA=120$\arcdeg~to $160$\arcdeg (red curve), and an average of the two (black curve). The average surrounding sky intensity has been subtracted from the cut.}
\label{RXBooCuts}
\end{figure}

We have made radial intensity cuts of the FUV emission around RX\,Boo, averaged over azimuthal wedges with their apexes centered on the star, and spanning the range from $PA=150$\arcdeg~to $190$\arcdeg~(around SymAx\,1) or $PA=120$\arcdeg~to $160$\arcdeg (around SymAx\,2) (Fig.\,\ref{RXBooCuts}) -- these wedges avoid Regions A and B, which result from imperfectly-subtracted bright point sources. Both cuts show a broad hump centered at $r\sim320{''}$. In the cut around SymAx\,1 the hump has a peak at $r\sim300{''}$, a local minimum, a more pronounced peak at $\sim350{''}$  and then a steady decline. In the cut around SymAx\,2, the hump shows a peak at $r\sim300{''}$ followed by a steady decline at $r\gtrsim340{''}$.  An average of the two cuts shows two small peaks at $r=300{''}$ and $r=350{''}$ followed by a steady decline. We take the average of the locations of the two peaks as an estimate of the termination shock radius, i.e., $R_1=325{''}$; the astrosheath outer radius is estimated to be $R_c=460{''}$.
 
\subsection{VX\,Eri}
\begin{figure}[htbp]
\begin{center}
\includegraphics[width=16cm]{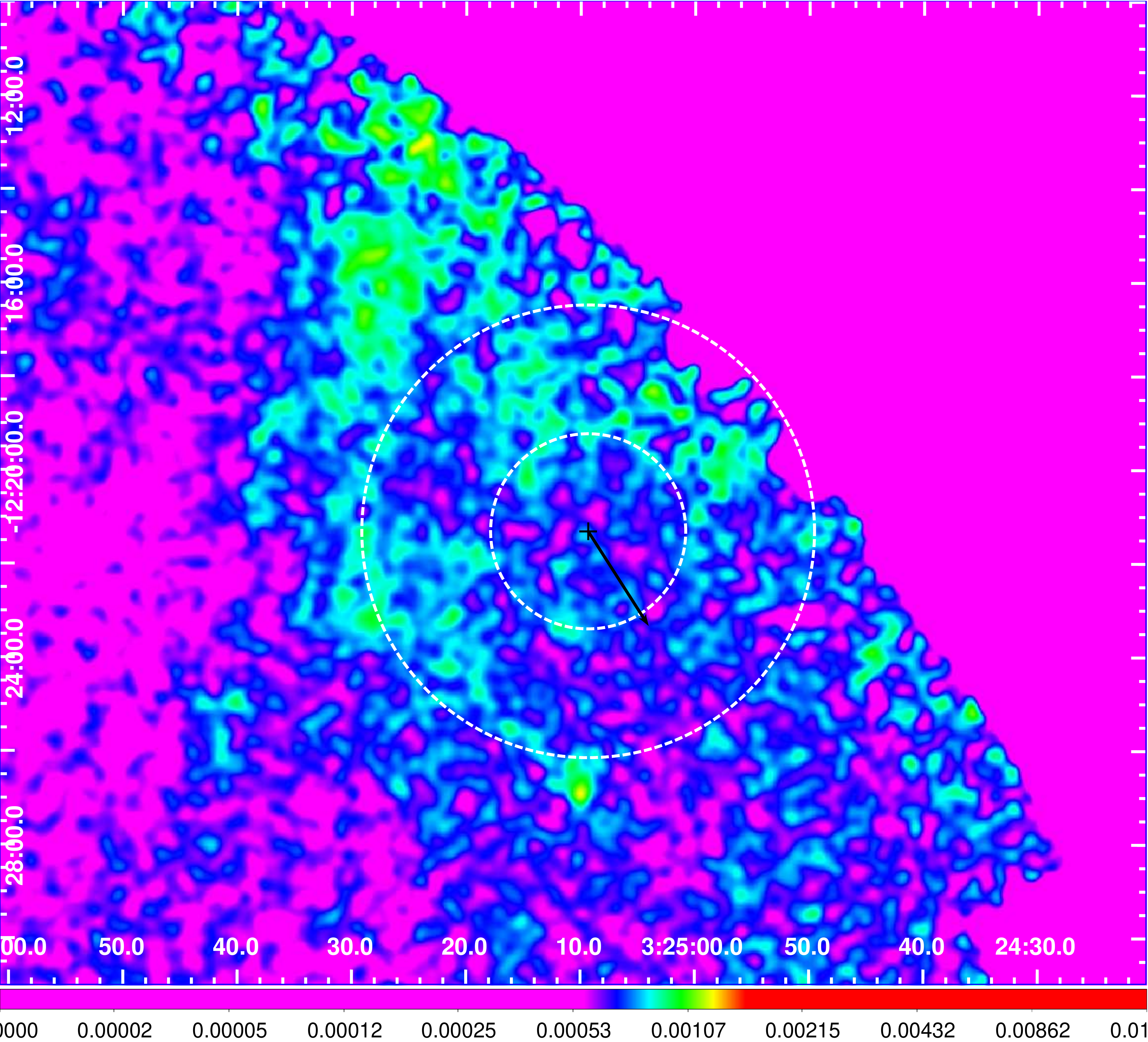}
\end{center}
\caption{The FUV emission towards VX\,Eri, imaged with GALEX. The dashed white circle (radius $290{''}$), delineates the radial extent of the termination-shock in the astrosphere, south-east of VX\,Eri. Black vector shows the star's proper motion of $14.4$\,mas\,yr$^{-1}$ at $PA=213\arcdeg$, magnified by a factor 10$^4$.  North is up and east is to the left.  The scalebar shows intensity in cps/pix.}
\label{VXEriFUV2}
\end{figure}

The FUV image of VX\,Eri (Fig.\,\ref{VXEriFUV2}) shows extended emission around the star. No such structure is seen in the NUV image. Unfortunately, VX\,Eri is located close to the edge of the FUV detector, so we are likely not seeing the full structure of the extended emission. The proper motion vector of VX\,Eri is directed towards $PA=-168.6\arcdeg$. Hence, the emission that is seen towards the north-east and north of the star, may represent part of the astrosphere's tail. In this case, we would expect the termination-shock intensity to peak in the opposite direction, i.e., towards the south and south-west, but we mostly see bright emission in the south-west quadrant. We therefore consider the detection of an astrosphere around VX\,Eri as tentative.
We have made a radial intensity of the FUV emission around VX\,Eri, averaged over an azimuthal wedge with its apex centered on the star, covering the azimuthal range $PA=100\arcdeg$ to $180\arcdeg$ (Fig.\,\ref{VXEriFUV2}). The cut shows a broad peak at $r\sim280{''}$, followed by a decrease in intensity at $r\sim290{''}$, then a local minimum, and then a second peak at $r=300{''}$, with a final steady decline to a plateau region starting at $r\sim460{''}$. The background sky intensity is reached at $r\sim580{''}$. We take the average of the two peaks near the edge of the bright UV region as a rough estimate of the radius of the termination shock,  $R_1=290{''}$; the astrosheath outer radius is estimated to be $R_c=460{''}$. The plateau region in the range $r\sim475{''}-580{''}$ may represent emission from swept-up ISM material between the outer edge of the astropause and the bow-shock interface separating the shocked and unshocked ISM.

\section{Astrospheres and Mass Loss}\label{astros}
Out of the 92 objects with long-exposure ($>$700\,s) FUV images examined for our survey, we found 10 objects with extended FUV emission. In these objects, with the exception of V\,Hya, 
we find the typical shock morphology expected for a spherical wind moving relative to, moving relative to, and interacting with, the ISM, in all objects (tentatively in VX\,Eri). For the latter, we find evidence of (i) its large parabolic outflows interacting with the ISM, and (ii) its collimated very hgh velocity outflows interacting with the circumstellar medium. We exclude it from further discussion and the astrosphere analysis (below) because the wind-ISM interaction in it cannot be analysed using the equations that govern a spherical wind-ISM interaction.

\begin{figure}[htbp]
\begin{center}
\includegraphics[width=16cm]{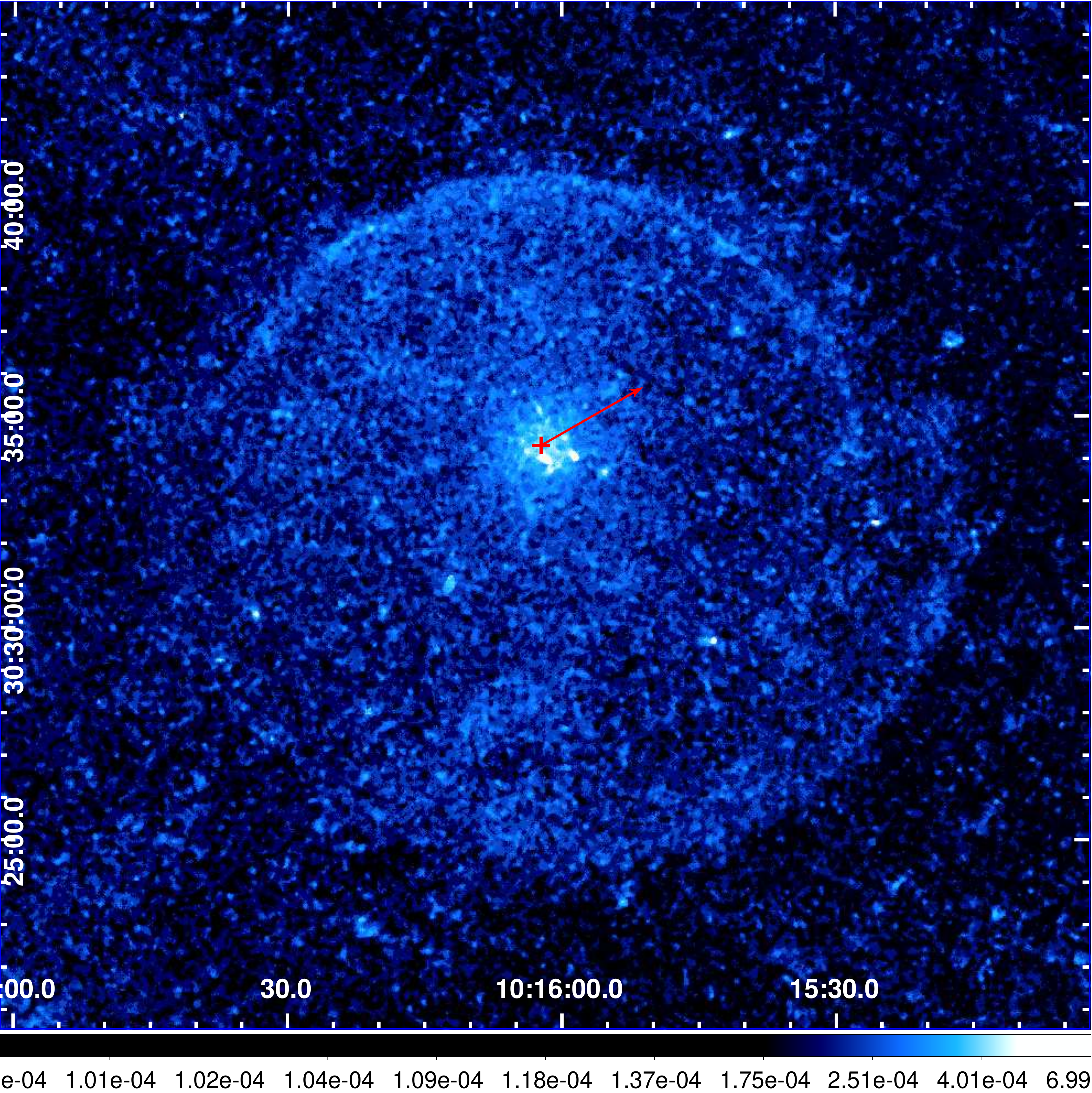}
\end{center}
\caption{The FUV emission towards CIT\,6, imaged with GALEX (adapted from \cite{Sahai2014}). Red vector shows the star's proper motion of $16.6$\,mas\,yr$^{-1}$ at $PA=300\arcdeg$, magnified by a factor 10$^4$; red cross shows the star's location. North is up and east is to the left. The scalebar shows intensity in cps/pix.}
\label{cit6-fd}
\end{figure}

Excluding V\,Hya, eight out of the nine objects have relatively large proper motions, and we find, as expected that the termination-shock region lies in a hemisphere that contains the proper motion vector. For 5 of these 8 objects (EY\,Hya, RT\,Vir, R\,Hya, RW\,Boo, and RX\,Boo) the symmetry axis of the termination-shock lies within $\lesssim\pm30\arcdeg$ of the proper motion vector. 
Of the remaining three (U\,Ant, W\,Hya and VX\,Eri), we don't have a full image of the astrosphere for VX\,Eri. For U\,Ant, the symmetry axis is quite uncertain (see \S\,\ref{uant-desc}, so the angle between the proper motion vector and the former could be anywhere in the range $20\arcdeg-35\arcdeg$. W\,Hya is thus the only source with a large proper motion, where it is clear that the termination shock is brightest in an azimuthal wedge offset by more than $\sim60\arcdeg$ from the proper motion vector. This suggests that the ambient ISM material around W\,Hya has a significant streaming velocity directed roughly westwards; alternatively, our inference about its astrosphere is incorrect -- we discuss this in more detail below (\S\,\ref{mejecta}). We note that like W\,Hya, the astrosphere of CIT\,6 also displays a relatively large azimuthal offset between the proper motion vector ($PA=-60\arcdeg$)  (Fig.\,\ref{cit6-fd}) and the symmetry axis of the termination shock ($PA\sim2\arcdeg$, \citep{Sahai2014}).

The proper motion of R\,LMi is relatively small (2.4\,mas\,yr$^{-1}$) compared to the other objects  ($\sim12{''}-79{''}$, Table\,\ref{tbl:mdotetc}), consistent with the roughly circular shape of its astrosphere.

\subsection{Detection Statistics and FUV Extinction}\label{extinc}
The fraction of AGB stars with wind-ISM interactions detected via emission in the GALEX FUV band\footnote{including the previously detected AGB stars, Mira, IRC+10216, and CIT\,6} appears to be relatively low, $\sim0.14$, especially in comparison to far-IR observations, which is $\sim0.4$ based on the \cite{Cox2012} study. Although a full consideration of sample selection biases in our study and that of \cite{Cox2012} is outside the scope of this paper\footnote{these biases will be discussed in detail in a follow-up paper which will cover the full survey}, an obvious factor that makes it relatively more difficult to detect FUV emission than far-IR emission is interstellar extinction. This extinction depends both on the object's distance and the height above the Galactic plane, $z$. We have derived the broad-band extinction in the GALEX FUV band for all our survey sources as follows: (i) the visual extinction $A_{V}$ for each source was computed using the the GALExtin (http://www.galextin.org) tool (\cite{Amores2021}), which estimates interstellar extinction based on both available 3D models/maps and distance \cite{Amores2005}, (ii) the FUV extinction, $A_{FUV}$, was computed using the values of $A_{FUV}$/$E_{B-V}$ and $A_{V}$/$E_{B-V}$ for the Milky Way in Table\,2 of \cite{Bianchi2011}. The distribution of $A_{FUV}$ for the undetected objects peaks at $\sim0.25$  (excluding a few outliers), with a full-width at half-maximum of 0.25, thus not significantly different from that for our small sample of detected objects. Hence it is unlikely that FUV extinction is the 
primary reason for the non-detection of FUV emission in the majority of the survey sample. More likely, other parameters that affect the presence and brightness of the wind-ISM interaction may be responsible such as the past history of mass-loss on long timescales ($\sim10^5$\,yr), the relative motion between the star and the ISM, and the local ambient density of the ISM (see below).

\subsection{Analysis of the Wind-ISM Interaction}\label{analy}
We now use our estimates of the radius of the termination shock, and the thickness of the astrosheath (listed in Table\,\ref{tbl:mdotetc}), for each of our targets where we can identify such structures, to derive various parameters characterising the mass-ejection history in these objects, following the methodology described by \cite{Sahai2014}. These parameters, listed in Table\,\ref{tbl:bows}, include the relative velocity between the star and the ISM (Col.\,14), the duration of, and total mass in, the unshocked wind (Cols.\,15--16), the duration of, and total mass in, the shocked wind (Cols.\,15--16), and the total ejected mass (Col.\,19), together with additional parameters relevant for the derivation of these parameters. We have adopted distances to our targets from \cite{Andriantsaralaza2022}, and when those are not available, from \cite{BailerJones2021} (Col.\,2, Table\,\ref{tbl:bows}). The mass-loss rates in Table\,\ref{tbl:mdotetc} have been scaled to (i) the adopted distances, and to (ii) a common value of the CO-to-H$_2$ abundance ratio -- $3\times10^{-4}$ for O-rich stars, and $10^{-3}$ for C-rich stars (Col.\,12, Table\,\ref{tbl:bows}). The largest uncertainty in the analysis described below is due to (i) uncertainties in the mass-loss rates, and (ii) assuming that the mass-loss rates and expansion velocties are constant with time.

In brief, we estimate the star's relative velocity $V_*$ through the surrounding ISM using the relationship between $l_1$, the distance of the termination shock from the star along the astropause's symmetry axis (i.e., the termination-shock standoff distance), and ${\rm V}_{*}$ using Eqn. 1 in \cite{VanBuren1988}. We reproduce this equation below for convenience, as given in \cite{Sahai2014}, who made a correction to the original equation (missing minus sign in the exponent of $\bar{\mu}_H$ in Eqn.\,1 of \cite{VanBuren1988}):
\begin{equation} l_1 (\text{cm}) =1.74\times10^{19}\,(\dot{M}_{*,-6} V_{w,8})^{1/2}\
(\bar{\mu}_H\,n_{\text{ISM}})^{-1/2}\,V_{*,6}^{-1}
\label{vanburen} 
\end{equation}
where $\dot{M}_{*,-6}$ is the stellar mass-loss rate in units of $10^{-6}$\,\my, $V_{w,8}$ is the wind velocity in units of $10^3$\,\kms, $\bar{\mu}_H$ is the dimensionless mean molecular mass per H atom, and $n_{\text{ISM}}$ is the ISM number density in cm$^{-3}$.

We make the simplifying assumption that the astropause's symmetry axis lies in the sky-plane, i.e., the inclination angle, $\phi=90^{\circ}$, hence $l_1=R_1$. We estimate the density of atomic and ionized hydrogen near each object (Cols\,8--9, Table\,\ref{tbl:bows}) based on its location in the Galaxy (i.e., Galactocentric radius R and height $z$ above the Galactic plane, Cols.\,5--6,  Table\,\ref{tbl:bows}) as described by \cite{Sahai2014} using published models (\cite{Kalberla2009, Gaensler2008}), and sum them to estimate the total ISM density. The duration (${\rm P}_w$) of, and total mass (${\rm M}_w$) in, the unshocked wind (i.e., within $r\le{\rm R}_1$) is estimated from the mass-loss rate and expansion velocity of the unshocked wind. The duration of mass-loss corresponding to the shocked wind in the astropause (${\rm R}_1<r<{\rm R}_c$), P$_s$, is estimated using its width (${\rm R}_c-{\rm R}$) and an estimate of the expansion velocity in this region, V$_s$, assuming that the shock is adiabatic (see \cite{Sahai2014} for details). The values for P$_s$ are likely to be lower limits for objects in which the astrosheath width, as a fraction of the termination-shock radius, (${\rm R}_c$-${\rm R}_1$)/${\rm R}_1$, is significantly less that $0.47$, expected for the adiabatic case (see Eqn. 2 of \cite{VanBuren1988}).

\subsubsection{Velocity of Stars relative to the local ISM}\label{vstar}
We compare the values of the relative velocities between the stars and their local ISM, V$_{*}$ (derived from our analysis above) to their peculiar space velocities, V$_t$ (derived using their local-standard-of-rest (LSR) radial velocities from CO mm-wave line observations, tangential proper motions from $GAIA$ DR3, and distances.) We have corrected V$_t$ for solar motion $v_{\odot}$ [$(U,V,W)=(8.5,13.38,6.49)$\,\kms] \citep{Coskunoglu2011}. We also assume, as in \citep{Cox2012}, that the local ISM for each star is at rest relative to the LSR -- this is a simplification for cases where the ISM may have a significant flow relative to the LSR, e.g., due to the presence of expanding super-bubbles (e.g., as found for $\alpha$\,Ori by \cite{Ueta2008b}).

We find that for 5 objects, EY\,Hya, R\,LMi, U\,Ant, RT\,Vir, and R\,Hya, there is reasonable agreement between V$_{*}$ and V$_t$ (i.e., within a factor $\lesssim2$), considering (i) the uncertainties in the mass-loss rates based on CO data, \mdot$_{w,0}$, which are easily a factor few (due to uncertanties in a variety of factors, such as the CO-to-H$_2$ abundance ratio and the outer radius of the CO-emitting circumstellar envelope, together with the use of empirical formulae or models that do not properly account for heating-cooling processes that determine the radial kinetic temperature distribution \cite{Sahai1990}), and (ii) intrinsic variations in the mass-loss rate over the very long periods (typically $\sim10^5$\,yr, see below) probed by the UV data. However, the discrepancies between V$_{*}$ and V$_t$ for the other 4 objects -- VX\,Eri, W\,Hya, RW\,Boo and RX\,Boo --  are much larger, although we may exclude VX\,Eri in this comparison, since we classify it's astrosphere as a tentative detection, and its gas mass-loss rate (expansion velocity) are unknown and estimated from the dust-mass loss rate (with an assumed gas-to-dust ratio). Since V$_{*}$ scales as (\mdot\,V$_e$)$^{1/2}$, and given that V$_e$ can be determined fairly reliably (typically to $\pm15$\% or better) from the CO line profiles, relatively large changes in \mdot~($\gtrsim10$) are required to make V$_{*}$ comparable to V$_t$. We note that for six out of nine stars with astrospheres, V$_{*}$ is less than V$_t$, suggesting that \mdot~was significantly higher in the past.

\subsubsection{Duration of Mass-Loss and Total Ejected Mass}\label{mejecta}
Our results show, that as for IRC+10216 and CIT\,6, the FUV emission traces the AGB stellar wind in our objects to much larger distances from the star, and therefore probes a much longer history of mass-loss, than typical mass-loss indicators such as CO mm-wave line emission which becomes undetectable, due to photodissociation by the interstellar UV field, at radii typically $\lesssim 2\times10^{17}$\,cm (or less) even for mass-loss rates as high as $\sim10^{-5}$\,\my, hence probing  mass ejection over a mere 6,500\,yr (assuming a typical expansion velocity of 10\,\kms). Far-IR observations can similarly probe the very extended mass-loss histories (e.g., \cite{Cox2012} and references therein) but these are sensitive to a minor component of the mass-ejecta, i.e., dust. We compare the FUV and far-IR results for the 6 objects that are common between our study and the far-IR studies (U\,Ant, R\,Hya, W\,Hya, RT\,Vir, V\,Hya and RX\,Boo). 

For two of these (U\,Ant, R\,Hya), we find that the wind-ISM interaction region lies at a significantly larger radius than could be seen in the far-IR data. In U\,Ant, the shell seen at $r\sim42{''}$ in he far-IR is due to a wind-wind interaction; we find, from the FUV data, a wind-ISM interaction region at $r\sim175{''}$. In R\,Hya, a shell at $r=93{''}$ is seen in the far-IR; we find, from the FUV data, a wind-ISM interaction region at $r\sim120{''}$. 

In W\,Hya, a shell at $r=93{''}$ is seen in the 70 and 160\,\micron~imaging, and a more distant one at $r\sim230{''}$ only at 160\,\micron, the latter corresponds to the radius of the termination shock that we find from the FUV data. In RT\,Vir, \cite{Maercker2022} derive a termination-shock radius of $85{''}$ from the far-IR imaging\footnote{although \cite{Cox2012} detected the wind-ISM interaction for this object, they did not provide a value for the termination-shock radius}, whereas the FUV data shows it to be smaller ($65{''}$).

In V\,Hya and RX\,Boo, \cite{Cox2012} did not find any emission related to a wind-wind or wind-ISM interaction.

We find that the duration of and total mass in the shocked wind (${\rm P}_s$, ${\rm M}_s$) are significantly larger than their corresponding values (${\rm P}_w$, ${\rm M}_w$) for the unshocked wind for all 9 objects that show a wind-ISM interaction -- this result is qualitatively consistent with that derived in the study by \cite{Maercker2022}, who have modeled the far-IR data from \cite{Cox2012} to derive dust masses in the interaction regions, and made estimates of the crossing time (time taken for material to reach the interaction region, therefore equivalent to ${\rm P}_w$) and build-up times (time taken by stellar wind to build up the mass in the interaction region, therefore equivalent to ${\rm P}_s$). Comparing results for the 3 objects with wind-ISM interaction that are common between \cite{Maercker2022} and our study, we find that P$_s$/P$_w$ for RT\,Vir, R\,Hya, and W\,Hya, is 5.5, 2.5, and 3.8, compared to 2.9, 2.6, and 12.5, respectively, from \cite{Maercker2022}\footnote{we exclude U\,Ant in this comparison, since it does not show the wind-ISM interaction in the far-IR}.

For five out of nine objects (EY\,Hya, R\,LMi, R\,Hya, W\,Hya, and RX\,Boo), the values of (${\rm R}_c$-${\rm R}_1$)/${\rm R}_1$ are significantly less than 0.47 (Table\,\ref{tbl:mdotetc}), showing that the shocked gas has cooled down significantly -- hence, V$_s$ is less than the adiabatic value, and the value of P$_s$, and consequently, M$_s$ and M$_t$ for these objects (Table\,\ref{tbl:bows}) are lower limits.

It is instructive to compare the total ejecta masses (M$_t$) that we derive for our sample of AGB stars to the mass that needs to be ejected before an AGB star begins its post-AGB journey, which is $\gtrsim$0.5\,\ms~for stars with MS progenitor masses of $\gtrsim1$\,\ms~(\cite{MillerBertolami2016}). We find that for all objects, except U\,Ant, M$_t<<0.5$\,\ms. This suggests that these stars are still far from the end of their AGB lifetimes. Even if the mass-loss rates were significantly higher in the past for the 5 stars where we find V$_t>$V$_*$ (i.e., EY\,Hya, R\,Hya, W\,Hya, RW\,Boo, and RX\,Boo; we exclude VX\,Eri in this consideration for reasons above, see \S\,\ref{vstar}), and we scale up the mass-loss rate for these in order to make V$_*$=V$_t$, the scaled-up values of M$_t$ are still significantly below 0.5\,\ms. Although for 4/5 of these stars, the value of M$_t$ is likely a lower limit because the actual value of the shock velocity V$_s$ is likely lower than the one derived assuming no cooling (see \S\,\ref{analy}), only for one of these (EY\,Hya) is the scaled-up value of M$_t$ high enough (0.17\,\ms) that a further increase resulting from a plausible decrease in V$_s$ due to cooling could raise it to $\sim0.5$\,\ms. 

The remaining 3 stars (R\,LMi, U\,Ant, and RT\,Vir) have V$_t<$V$_*$, hence the mass-loss rate, and thus M$_t$ would have to be lower, in order for V$_*$=V$_t$. U\,Ant is a carbon star, and believed to have descended from a progenitor of mass 3.6\,\ms, slightly above the average of the range of progenitor masses of carbon stars (2.5--4\,\ms) (\cite{Kastner2021}), and so even though for it, M$_t\sim1$\,\ms, it is also not close to transitioning to the post-AGB phase. Amongst the other two C stars with previously detected astrospheres in the UV, IRC+10216 and CIT\,6, only IRC+10216 shows a large enough ejecta mass ($\sim1.4$\,\ms: \cite{Sahai2010}) that it can be considered very close to transitioning to the post-AGB evolutionary phase.

The morphology of the FUV emission towards W\,Hya is remarkable, showing an unprecendented variety of azimuthal and radial structures, and indicating a fairly complex mass-loss history that includes multiple spherical and collimated outflows. The presence of collimated outflows is surprising, considering that these generally manifest themselves during the very late AGB or the early post-AGB phase, i.e., the pre-planetary nebula phase (e.g., \cite{Sahai1998,Sahai2011}), whereas, based on the small inferred value of ejected mass in it, W\,Hya is still very far from the end of its AGB lifetime. A possible solution to this predicament is that W\,Hya is actually near the end of its AGB lifetime, and that its wind-ISM interaction region lies at a much larger radius than inferred from our current analysis, but is too faint to be seen. In this scenario, R$_1$ (and thus $l_1$) would be larger, and the past mass-loss rate would have to be much higher in order to reconcile the value of V$_*$ with V$_t$ -- e.g., with R$_1$ (and R$_c$) a factor 2 larger, the value of M$_t$ would be $\sim0.55$\,\ms~(or even higher if V$_s$ was lower than its adiabatic value.) 

\subsection{Comparison with Theory}
Our FUV imaging enables us to detect an important component of the wind-ISM interaction that is expected from theory, but cannot be distinguished in the far-IR studies of astrospheres -- namely, the region between the outer edge of the astropause and the bow-shock interface separating the shocked and unshocked ISM. As noted earlier, in 5 objects (VX\,Eri, EY\,Hya, U\,Ant, RT\,Vir, VX\,Eri) we can see a faint FUV emission ``plateau" from this region in the radial intensity cuts, that lies just beyond the region of steeply declining intensity that characterises the astrosheath. This faint emission likely arises in the shocked ISM. Such bow-shock emission has also been found in the astrospheres of IRC+10216 \citep{Sahai2010} and CIT\,6 \citep{Sahai2014}. 

We note that while the FUV emission from some of the astrospheres found in our study shows the expected limb-brightened appearance as seen in IRC+10216 and CIT\,6 (e.g., R\,LMi, U\,Ant, RX\,Boo, and possibly VX\,Eri), the others show a ``filled" appearance, i.e., the average brightness of the emission interior to the termination-shock is comparable to that at the shock (EY\,Hya), or in fact rises inwards (RT\,Vir, R\,Hya, and W\,Hya)\footnote{while making this inference, we have disregarded the sharp central peak that can be seen in some objects}. The `filled" objects may result from various kinds of hydrodynamic fluid instabilities revealed in numerical simulations of the wind-ISM interaction.

The FUV data on wind-ISM interactions presented in our study can be very useful for providing strong constraints on numerical hydrodynamical simulations of such interactions. Many such studies have been undertaken in the past with pioneering work by \cite{Blondin98} and \cite{Comeron98}, in which the morphology and dynamical evolution of wind bow shocks produced by runaway stars in the diffuse ISM was analysed. \cite{Blondin98} examined the instabilities of isothermal stellar wind bowshocks and concluded that ragged, clumpy bowshocks should be expected to surround stars with a slow, dense wind, which is moving through the ISM with a Mach number greater than a few (i.e., with relative velocity of order $60$\,\kms). \cite{Comeron98} found a diversity of structures, even with moderate changes in basic input parameters: depending on these, the bowshocks may have a simple or layered structure, or may not even form at all.

In the case of AGB astrospheres, such simulations can be used to explore the effects on the astrospheres of varying physical parameters of the stars (mass-loss rate, wind-velocity, velocity relative to the ISM, including time-variations) and the ambient ISM (density, temperature) systematically on the detailed shape and structure of the astrosphere in both gas and dust. The combination of the FUV imaging presented here, together with the far-IR imaging, provides a unique database for the study of wind-wind and wind-ISM interactions.

A unique strength of simulations is that they enable one to utilise the information present in the microstructure of the shocks. Such structure can result from different kinds of fluid instabilities, such as Rayleigh-Taylor (RT) and Kelvin-Helmholtz (KH) instabilities. RT instabilities can be quenched by a magnetic field, thus the presence and/or absence of RT fingers can be used to set constraints on the presence of a magnetic field. The KH instability together with the RT instability can form mushroom-shaped structures on the ends of RT fingers. KH time-dependent turbulent eddies can become large enough to affect the large-scale morphology of the shocked gas. Other instabilities include the non-linear thin shell instability \citep{Vishniac94}, the transverse acceleration instability \citep{Dgani96}, and large-scale vortex instabilities \citep{Wareing2007}.

The above simulations can be made more useful by including relevant physical processes such as heating and cooling. Treating the gas and dust as separate fluids, and accounting for dust charge/size distribution and destruction is especially important for modeling the far-IR data. \cite{Cox2012} have carried out 7 simulations (without dust radiative cooling, charge or destruction) in order to determine how the morphology of the bow shock varies with various stellar wind and ISM properties. \cite{Villaver12} present another set of pure hydro simulations, but include wind modulations prescribed by stellar evolution calculations, and cover a range of expected relative velocities ($10-100$\,\kms), ISM densities $0.01-1$\,cm$^{-3}$), and stellar progenitor masses (1 and 3.5\,\ms). Such studies are most useful when accompanied by observational predictions, e.g., the hydrodynamical modeling of $\alpha$\,Ori's astrosphere by \cite{mohamed12}. 

\section{The H$_2$ line spectrum of an astrosphere and future spectroscopic observations}\label{shs}
One of the major limitations of the FUV data as a probe of wind-wind or wind-ISM interactions is that no spectra of the FUV emission are available to confirm the current hypothesis for the nature and origin of the emission\footnote{the far-IR data suffer from a similar limitation, since the [OI]63\,\micron~and [CII]158\,\micron~line emission may contribute to the emission seen in the broad-band PACS 70 and 160\,\micron~filters used in the far-IR observations by \cite{Cox2012}} -- collisional excitation of the H$_2$ Lyman-Werner band line emission by hot electrons in shocked gas. This hypohesis is based on modeling of very low-resolution grism spectroscopy of the FUV emission from Mira's astrosphere by \cite{Martin2007} -- no spectra showing actual lines could apparently be extracted against the diffuse FUV background\footnote{The spectral fits shown in their Figure S3 are model fits to the spatial intensity in the grism image}. A full understanding of the nature of FUV emission from astrospheres requires high-resolution (R$\sim$100,000) spectroscopic observations.

\begin{figure}[htbp]
\begin{center}
\includegraphics[width=10cm]{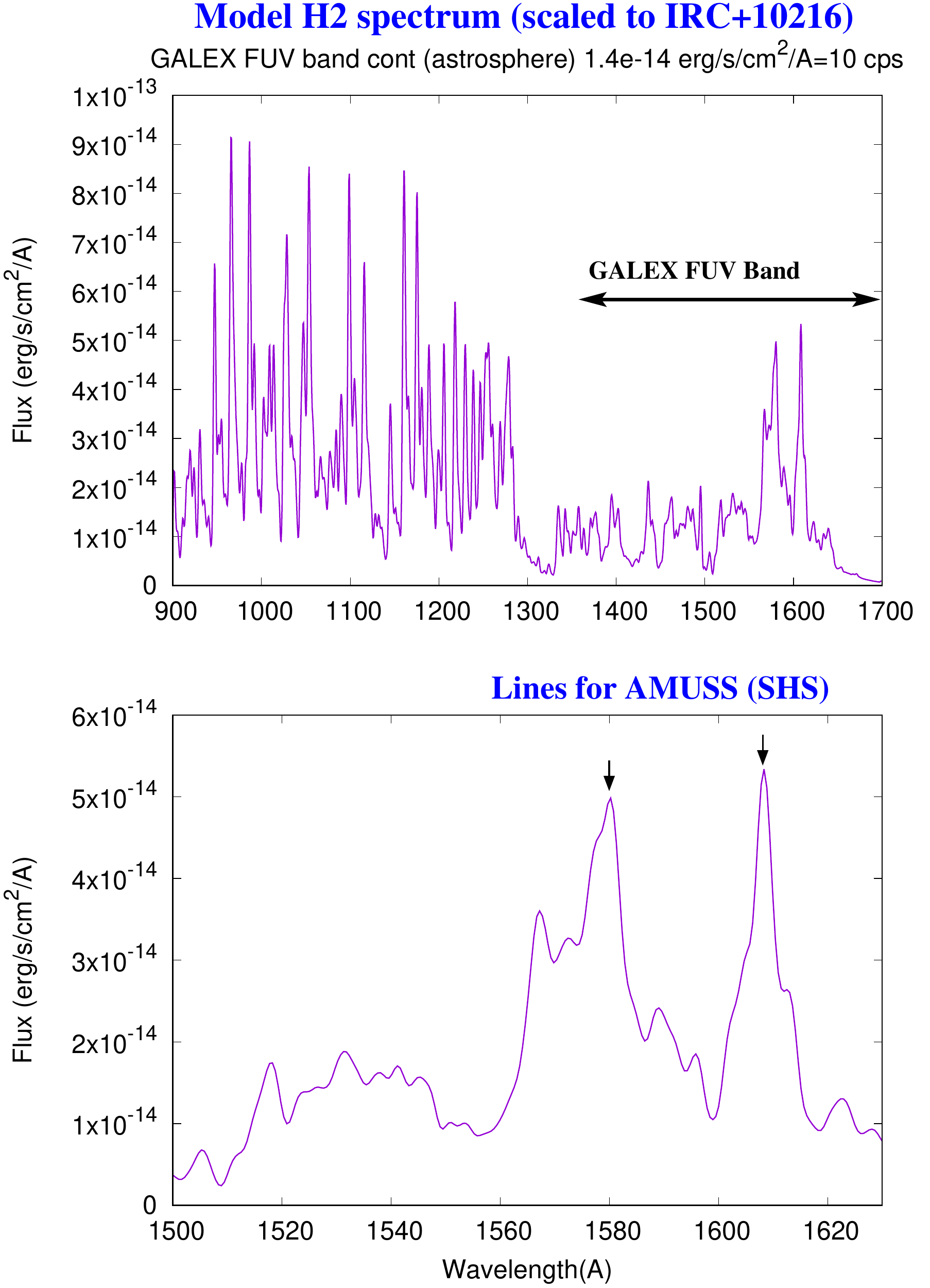}
\end{center}
\caption{A model FUV-band spectrum due to electron impact excitation of H$_2$, scaled to fit the FUV emission flux observed in an elliptical patch of size $210{''}\times110{''}$ covering the brightest part of the leading edge of the astrosphere of IRC+10216. {\it Top} panel shows the full spectrum of Lyman-Werner band lines; whereas the {\it bottom} panel shows an expanded view of the spectrum indicating the centers of two wavelength windows that would be optimum for observations with AMUSS.
}
\vskip -0.15in
\label{h2mods}
\end{figure}

We use a simple model of the FUV emission resulting from electron impact excitation of H$_2$, in order to estimate the expected spectrum from an astrosphere. We generate a model spectrum using a code written by Dr. X. Liu (Space Environment Technologies)\footnote{code has been updated in 2019} \citep{Liu96a,Liu96b}, for electron impact energy 100 eV (the spectrum is relatively insensitive to this energy) and H$_2$ rotational temperature 300\,K. This spectrum is then convolved with the GALEX FUV band response and scaled in order to reproduce the broad-band GALEX FUV flux emitted by an astrosphere (in \fluxu~or cps, where 1 cps = $1.4\times10^{-15}$\,\fluxu). 
Using this model, we compute the spectrum of the FUV emission from the astrosphere of IRC+10216. We extracted the total flux (10\,cps) from an elliptical patch of size $210\times110$\,pix$^2$ ($A_{shs}$) covering the brightest part of the leading edge of the astrosphere of IRC+10216, and then 
obtained a prediction of the astrosphere spectrum (Fig.\,\ref{h2mods}) as described above. 

The Cyclical Spatial Heterodyne Spectrometer (SHS) can provide the sensitivity required to obtain high-resolution spectra of this faint line emission from IRC+10216's astrosphere in particular, and the astrospheres of AGB stars in general -- a sensitivity that traditional slit spectrographs lack. This interferometric instrumental technique, conceived at the University of Wisconsin in the 1990's, and built and tested in the laboratory at visible and UV wavelengths, demonstrated concept-feasibility and performance characteristics \citep{Harlander1992}. Radial velocity resolved detections of interstellar [OII]3727 line emission were obtained by \citep{Mierkiewicz2006,Mierkiewicz2007} from the University of Wisconsin’s Pine Bluff Observatory using an SHS. Payload performance at 1550\,\AA~of an SHS onboard a sounding rocket has also been tested (\cite{Watchorn2010}).

SHS employs a miniature all-reflective two-beam interferometric technique to obtain spectra at a very high spectral resolving power (R$\sim100,000$) of light collected from a large diffuse, faint emission region (\cite{Dawson2009,Hosseini2012,Hosseini2019,Hosseini2020}). Field testing carried out by \cite{Corliss2015} has demonstrated that reflective SHS instruments can deliver effective interferometric performance in the visible to far-ultraviolet wavelengths with commercial optics of moderate surface quality.
The AMUSS (Astrophysics Miniaturized UV Spatial Spectrometer) concept, submitted in response to RFI NNH17ZDA010L: Possible NASA Astrophysics SmallSats, utilises such an instrument (with mass $<5.4$\,kg for 3 different UV lines served by 3 SHS instruments) that when coupled to a small-aperture ($\lesssim30$\,cm) space-based telescope, can provide high-resolution spectra of diagnostic UV lines emitted by extended, faint astrophysical objects, such as astrospheres, the ISM in our galaxy, and the circumgalactic medium in nearby galaxies (\cite{HosseiniSahai19}). 

The lower panel of Fig.\,\ref{h2mods} shows an expanded view of the spectrum of IRC+10216's astrosphere, with black arrows indicating the centers of two wavelength windows covering lines that would be optimum for observations with AMUSS. Although there are a plethora of strong lines at shorter wavelengths ($\lesssim1300$\,\AA), AMUSS would be less sensitive to them because of the steep decrease in the quantum efficiency of UV detectors at shorter wavelengths. \cite{HosseiniSahai19} find that AMUSS, coupled to a 30\,cm space-based telescope, can obtain a spectrum (with 10\,\kms~resolution) in a limited bandpass\footnote{since the noise in a spectrum obtained with AMUSS varies as the square-root of the total bandpass, it is advantageous to limit the bandpass as much as possible, while still being able to detect the line(s) of interest} of (say) $\sim(1600-1620)$\,\AA, of an astrosphere that is as faint as the faintest source in our survey, RW\,Boo, with about 5\,hr of integration time per source, and detect the 1610\,\AA~line with $\gtrsim5\sigma$. RW\,Boo has a flux of $1$\,cps, i.e., a factor $\sim$10 less than IRC+10216 in an elliptical aperture with area $A_{shs}$ covering the brightest part of its astrosphere. For comparison, the brightest source in our survey, U\,Ant, has a flux of 5.5 cps in an elliptical aperture with area $A_{shs}$, covering the brightest part of its astrosphere.

\section{Conclusions}\label{sumry}
Using the GALEX archive, we have discovered extended structures around a small sample of AGB stars emitting in the far-ultraviolet (FUV) band. 
\begin{enumerate}
\item In nine out of ten objects, we find the typical morphology expected for a spherical wind moving relative to, and interacting with the ISM to produce an astrosphere (including one  tentative detection). The exception is the carbon star, V\,Hya, whose mass-ejection is known to be highly aspherical; in it we find evidence of its large parabolic high-velocity outflows interacting with the ISM, and its collimated, extreme velocity outflows interacting with the circumstellar medium.
\item W\,Hya shows a complex morphology with multiple azimuthal and radial structures, in addition to its astrosphere.
\item For eight objects with relatively large proper motions, we find (as expected) that the termination-shock region lies in a hemisphere that contains the proper motion vector. For five out of eight objects (EY\,Hya, RT\,Vir, R\,Hya, RW\,Boo, and RX\,Boo) the symmetry axis of the termination-shock lies within $\lesssim\pm30\arcdeg$ of the proper motion vector. One object (R\,LMi), which has a very small proper motion by far, compared to the others, shows a roughly circular morphology for its astrosphere.
\item Radial intensity cuts for each source, averaged over large azimuthal wedges locate the termination shock, the astropause and its outer edge. In a few objects, the cuts also reveal faint emission just outside the astropause that likely arises in shocked ISM material. 
\item The radii of the termination shock and the width of the astropause deribed from the intensity cuts, together with published mass-loss rates and wind expansion velocities, have been used to determine the total mass lost and mass loss duration for each source -- we find that the duration of, and total mass in, the shocked wind are significantly larger than their corresponding values for the unshocked wind.
\item The total derived ejecta masses for all eight stars with well-detected astrospheres are small (or very small) fractions of the minimum mass that needs to be lost before such stars enter the post-AGB evolutionary phase.
\item The NUV images of these objects do not show the extended emission structures, indicating that the FUV emission is due to H$_2$ Lyman-Werner band lines within the wide GALEX FUV filter bandpass, excited by collisions with hot electrons produced as a result of the shock interaction, as has been hypothesized for other astrospheres detected in FUV emission. We derive a model spectrum of the FUV emission from a representative bright region in the astrosphere of IRC+10216, assuming it results from this mechanism.
\item We show that a Spatial Heterodyne Spectrometer instrument, mounted on a relatively small-aperture space-based telescope, can obtain high-velocity resolution spectra of the faint FUV emission from astrospheres around AGB stars in 5\,hr or less per source, in order to confirm (or refute) the origin of this emission as resulting from H$_2$ Lyman-Werner band lines.
\end{enumerate}

\begin{acknowledgments}
We thank an (anonymous) referee whose comprehensive review has helped us improve this paper. We thank G.\,Bryden (JPL) and M.\,Morris (UCLA) for discussions related to velocities and frames of reference.
RS's contribution to the research described in this publication was carried out at the Jet Propulsion Laboratory, California Institute of Technology, under a contract with NASA. RS thanks NASA for financial support via a GALEX GO and ADAP award. BS thanks JPL for a NASA Student Independent Research Internship (SIRI). This work has made use of data from the European Space Agency (ESA) mission Gaia (https://www.cosmos.esa.int/gaia), processed by the Gaia Data Processing and Analysis Consortium (DPAC, https://www.cosmos.esa.int/web/gaia/dpac/consortium). Funding for the DPAC has been provided by national institutions, in particular the institutions participating in the Gaia Multilateral Agreement.
\end{acknowledgments}.


\begin{table}
\caption{Observation Log}
\label{tbl:log}
\begin{tabular}{c c c c c c}
\hline\hline      
Name   & Sp.Typ. & FUV\,Image\,Root\tablenotemark{a} & FUV\,Exp.Time\tablenotemark{b} & NUV\,Image\,Root\tablenotemark{c} & NUV\,Exp.Time\tablenotemark{d} \\
       &         &                                   &  (s)                           &                                   & (s)    \\
\hline
VX Eri&  M3/4III&       MISWZS03\_27400\_0183\_0002&  722.55&  MISWZS03\_27400\_0183\_0002&  722.55 \\
EY Hya&  M7&            MISDR1\_24292\_0467\_0007&  3352.05&  MISDR1\_24292\_0467\_0007&  3352.05 \\
R LMi&   M6.5-9e&       GI1\_023004\_HIP47886\_0002&  1605.1&  GI1\_023004\_HIP47886\_0001&  3290.2 \\
U Ant&   C-N3&          GI5\_021008\_U\_Ant\_0002&   1119.05&  GI5\_021008\_U\_Ant\_0001&   6379.1 \\
V Hya&   C-N:6&         GI1\_023019\_HIP53085\_0001&  2696.2&  GI1\_023019\_HIP53085\_0001&  2696.2 \\
RT Vir&  M8III&         GI5\_021002\_RT\_Vir\_0001&  3473.85&  GI5\_021002\_RT\_Vir\_0001&  9079.9 \\
R Hya&   M6-9e&         GI5\_021003\_R\_Hya\_0001&   6029.75&  GI5\_021003\_R\_Hya\_0001&   7651.05 \\
W Hya&   M7.5-9e&       GI5\_021004\_W\_Hya\_0001&   4420.4&  GI5\_021004\_W\_Hya\_0001&   4420.4 \\
RW Boo&  M5III:&        GI1\_023006\_HIP71802\_0003&  1721.&  GI1\_023006\_HIP71802\_0001&  5113.0 \\
RX Boo&  M7.5-M8&       GI5\_021005\_RX\_Boo\_0001&  5161.45&  GI5\_021005\_RX\_Boo\_0001&  7657.65 \\
\hline
\end{tabular}
\tablenotetext{a}{OBJECT keyword in FUV Image fits file header}
\tablenotetext{b}{Total Exposure time for FUV image}
\tablenotetext{c}{OBJECT keyword in NUV Image fits file header}
\tablenotetext{b}{Total Exposure time for NUV image}
\end{table}

\begin{sidewaystable}
\caption{Stellar, Mass-Loss \& Measured Astrosphere Parameters}
\label{tbl:mdotetc}
\hspace*{-8cm}\begin{tabular}{c c c c c c c c c c c c c c l}
\hline\hline
Name                   & Long.  & Lat.   & PM               & D$_0$ & $\dot{\rm M}_{w,0}$ & V$_{lsr}$ &V$_e$  &f(CO)$_0$   & Ref.\tablenotemark{(d)} & Aver.Int.            & R$_1$  & R$_c$  & (R$_c$--R$_1$)/R$_1$ & Morph. \\
                       & (Deg.) & (Deg.) & (mas\,yr$^{-1}$) & (kpc) & (10$^{-6}$\,\my)    & (\kms)    &(\kms) & ($10^{-3}$)&                         & (10$^{-4}$\,cps/pix) & ({''}) & ({''}) &                      &      \\
(1)\tablenotemark{(a)} & (2)    & (3)    &  (4)             & (5)   &  (6)                &    (7)    & (8)   & (9)        & (10)                    & (11)                 & (12)   & (13)   & (14)                 & (15) \\
\hline
VX\,Eri& 198.3011& -51.1339& 14.79& 0.657& 0.023& -21.9& 10& 0.3& 1& 1.8& 290.0& 460.0& 0.59& w-ISM+w-w? \\
EY\,Hya& 225.3952& 26.1423& 11.55& 0.300& 0.25& 22.5& 11& 0.2& 2& 0.84& 130.0& 170.0& 0.31& w-ISM+bs \\
R\,LMi& 190.5954& 49.7711& 2.398& 0.330& 0.26& 0& 7.5& 0.3& 3& 0.34& 295.0& 375.0& 0.27& w-ISM \\
U\,Ant& 276.2241& 16.1419& 31.61& 0.260& 10& 24.5& 19& 1.0 & 4& 4.3& 175.0& 255.0& 0.46& w-ISM+bs+w-w? \\
V\,Hya& 268.9649& 33.6014& 16.31& 0.380& 2.5\tablenotemark{(b)}& -17.4 & 15,45,200\tablenotemark{(b)}& 1.0 & 7& 0.44  & 365\tablenotemark{(c)} & 430\tablenotemark{(c)} & 0.18\tablenotemark{(c)} & (see text, \S\,\ref{vhya-desc}) \\
RT\,Vir& 310.3571& 67.8959& 41.05& 0.136& 0.5& 17.4& 7.8& 0.2& 2& 0.57& 65.0& 95.0& 0.46& w-ISM+bs \\
R\,Hya& 314.2230& 38.7498& 55.48& 0.118& 0.16& -10& 12.5& 0.2& 5& 0.73& 120.0& 145.0& 0.21& w-ISM \\
W\,Hya& 318.0224& 32.8108& 79.01& 0.104& 0.078& 41& 8.5& 0.2& 5& 1.1& 220.0& 290.0& 0.32& w-ISM+w-w \\
RW\,Boo& 50.0855& 65.7340& 15.61& 0.307& 0.044& 5\tablenotemark{d}& 17.3& 0.3& 6& 0.39& 260.0& 310.0& 0.19& w-ISM+bs? \\
RX\,Boo& 34.2774& 69.2127& 52.52& 0.128& 0.0649& 2& 11.2& 0.3& 6& 0.66& 325.0& 475.0& 0.46& w-ISM \\
\hline

\end{tabular}
\tablenotetext{a}{Column headings: (1) Name, (2) Galactic Longitude, (3) Galactic Latitude, (4) Proper Motion, (5) Distance to star, (6) Mass-Loss Rate, (7) Stellar Velocity, relative to the LSR (8) Wind Expansion Velocity, (9) CO-to-H$_2$ abundance ratio, (10) Reference for Cols. 4--6, (11) Average FUV Intensity of Termination Shock (10$^{-4}$ counts per second (cps) pix$^{-1}$=$0.622\times10^{-19}$\,\intunit, (12) Termination Shock radius (angular) from radial intensity cut, (13)  Astropause radius (angular) from radial intensity cut, (14) Astrosheath Fractional Width (15) Morphology of extended FUV emission -- w-ISM\,(wind-ISM interaction),\,bs (emission from shocked ISM),\,w-w (wind-wind interaction)}
\tablenotetext{b}{The mass-outflow from V\,Hya is complex with multiple components; the mass-loss rate value is the sum of the mass-loss rates for the 3 components with different expansion velocities (listed in Col. 7) as identified by \cite{Knapp1997}}
\tablenotetext{c}{These values apply to the western-most part of the elliptical ring centered to the west of the star, extracted from a radial intensity cut spanning a narrow azimuthal wedge around the ring's minor axis.}
\tablenotetext{d}{The value of V$_{lsr}$ has been inferred from the CO J=1--0 and 2--1 line spectra in \cite{DiazLuis2019}, since the tabulated value in this reference (-4.99\,\kms) appears have an incorrect sign.}
\tablenotetext{d}{References: (1) this study, (2) \cite{Olofsson2002}, (3) \cite{Danilovich2015}, (4) \cite{Kerschbaum2017}, (5) \cite{DeBeck2010}, (6) \cite{DiazLuis2019}, (7) \cite{Knapp1997}}

\end{sidewaystable}


\clearpage
\begin{rotatetable}
\newgeometry{margin=-2cm} 
\voffset=0.1in
\headsep=3in
\begin{longtable}{p{1.2cm} p{0.7cm} p{0.8cm} p{1.45cm} p{0.8cm} p{0.9cm} p{0.8cm} p{0.8cm} p{0.8cm} p{0.8cm} p{0.9cm} p{0.8cm} p{0.7cm} p{0.7cm} p{1.0cm} p{0.9cm} p{0.9cm} p{1.0cm} p{0.9cm} p{0.8cm}}
\hline
\caption{Physical Properties of Astrospheres} \\
\hline 
Name                   & Dist. & $A_{FUV}$  & FUV\,Int.  &${\rm R}$   &  $z$  & $h_z$ & n(HI)       & n(H$_+$)    & R$_1$ & R$_c$-R$_1$ & $\dot{\rm M}$& V$_t$   & V$_*$  & P$_w$ & M$_w$  & V$_s$  & P$_s$ & M$_s$ & M$_t$ \\
(1)\tablenotemark{(1)} & (2)   & (3)     &(4)         & (5)        &  (6)  &  (7)  & (8)         & (9)         & (10)  & (11)        & (12)            & (13)   & (14)  & (15)   & (16)   & (17)  & (18)  & (19)  & (20) \\
\hline
\endhead
VX\,Eri& 0.657& 0.38& 1.6e-19& 8.722& -0.512& 0.156& 0.000& 0.122& 191& 112& 0.023& 51& 2.3& 90322& 0.0021& 0.83& 635368& 0.015& 0.017 \\
EY\,Hya& 0.422& 0.44& 8.1e-20& 8.600& 0.186& 0.154& 0.301& 0.234& 54.9& 16.9& 0.33& 32& 15& 23642& 0.0078& 0.92& 87295& 0.029& 0.037 \\
R\,LMi& 0.320& 0.53& 3.6e-20& 8.533& 0.244& 0.153& 0.145& 0.209& 94.4& 25.6& 0.244& 3.6& 7.7& 59668& 0.015& 0.62& 194174& 0.047& 0.062 \\
U\,Ant& 0.294& 0.36& 3.8e-19& 8.304& 0.082& 0.150& 0.738& 0.289& 51.5& 23.5& 12.8& 50& 95& 12837& 0.16& 1.58& 70420& 0.9& 1.1 \\
V\,Hya& 0.311& 0.34& 3.9e-20& 8.339& 0.172& 0.150& 0.361& 0.241& ... & ... & ... & ... & ... & ... & ... & ... & ... & ...  & ...  \\
RT\,Vir& 0.227& 0.19& 4.3e-20& 8.275& 0.210& 0.149& 0.231& 0.223& 14.8& 6.81& 0.929& 47& 86& 8968& 0.0083& 0.65& 49667& 0.046& 0.054 \\
R\,Hya& 0.126& 0.094& 5e-20& 8.262& 0.079& 0.149& 0.757& 0.290& 15.1& 3.15& 0.122& 35& 25& 5734& 0.0007& 1.04& 14335& 0.0017& 0.0024 \\
W\,Hya& 0.087& 0& 6.9e-20& 8.276& 0.047& 0.149& 0.854& 0.309& 19.1& 6.09& 0.0364& 52& 8.6& 10675& 0.00039& 0.71& 40758& 0.0015& 0.0019 \\
RW\,Boo& 0.253& 0.22& 3e-20& 8.264& 0.231& 0.149& 0.175& 0.214& 65.8& 12.7& 0.0299& 19& 5.6& 18025& 0.00054& 1.44& 41596& 0.0012& 0.0018 \\
RX\,Boo& 0.139& 0.088& 4.5e-20& 8.289& 0.130& 0.149& 0.540& 0.262& 45.2& 20.9& 0.0765& 35& 7.3& 19121& 0.0015& 0.93& 105901& 0.0081& 0.0096 \\
\hline

\label{tbl:bows}
\end{longtable}
\tablenotetext{1}{Column headings: (1) Name, (2) Adopted Distance to star (kpc) -- from \cite{Andriantsaralaza2022} for R\,LMi, U\,Ant, V\,Hya, RT\,Vir, R\,Hya, W\,Hya, \& RX\,Boo; from \cite{BailerJones2021} for VX\,Eri, EY\,Hya, \& RW\,Boo, (3) Interstellar Extinction in the GALEX FUV band,  (4) Average FUV intensity at Termination Shock (\intunit),  
(5) Galactocentric Radius (kpc), (6) Height above Galactic Plane (kpc), (7) Disk Scale Height of Atomic Hydrogen (kpc), (8) Atomic Hydrogen Density (cm$^{-3}$), (9) Ionized Hydrogen Density (cm$^{-3}$), (10) Termination Shock Distance ($10^3$ au), (11) Thickness of Astropause ($10^3$ au), (12) Mass-Loss Rate, scaled to distance in Col.\,4 (10$^{-6}$\,\my), (13) Star velocity from radial velocity and tangential proper motion (\kms), (14) Star velocity relative to local ISM (\kms), (15) Duration of unshocked wind (yr), (16) Mass of unshocked wind (\ms), (17) Expansion Velocity of shocked wind (\kms), (18) Duration of shocked wind (yr), (19) Mass of shocked wind (\ms), (20) Total Mass (unshocked+shocked) (\ms)
}
\end{rotatetable}
\restoregeometry

\clearpage
\clearpage

\clearpage

\clearpage

\clearpage

\clearpage

\clearpage

\begin{figure}[htbp]
\begin{center}
\includegraphics[width=16cm]{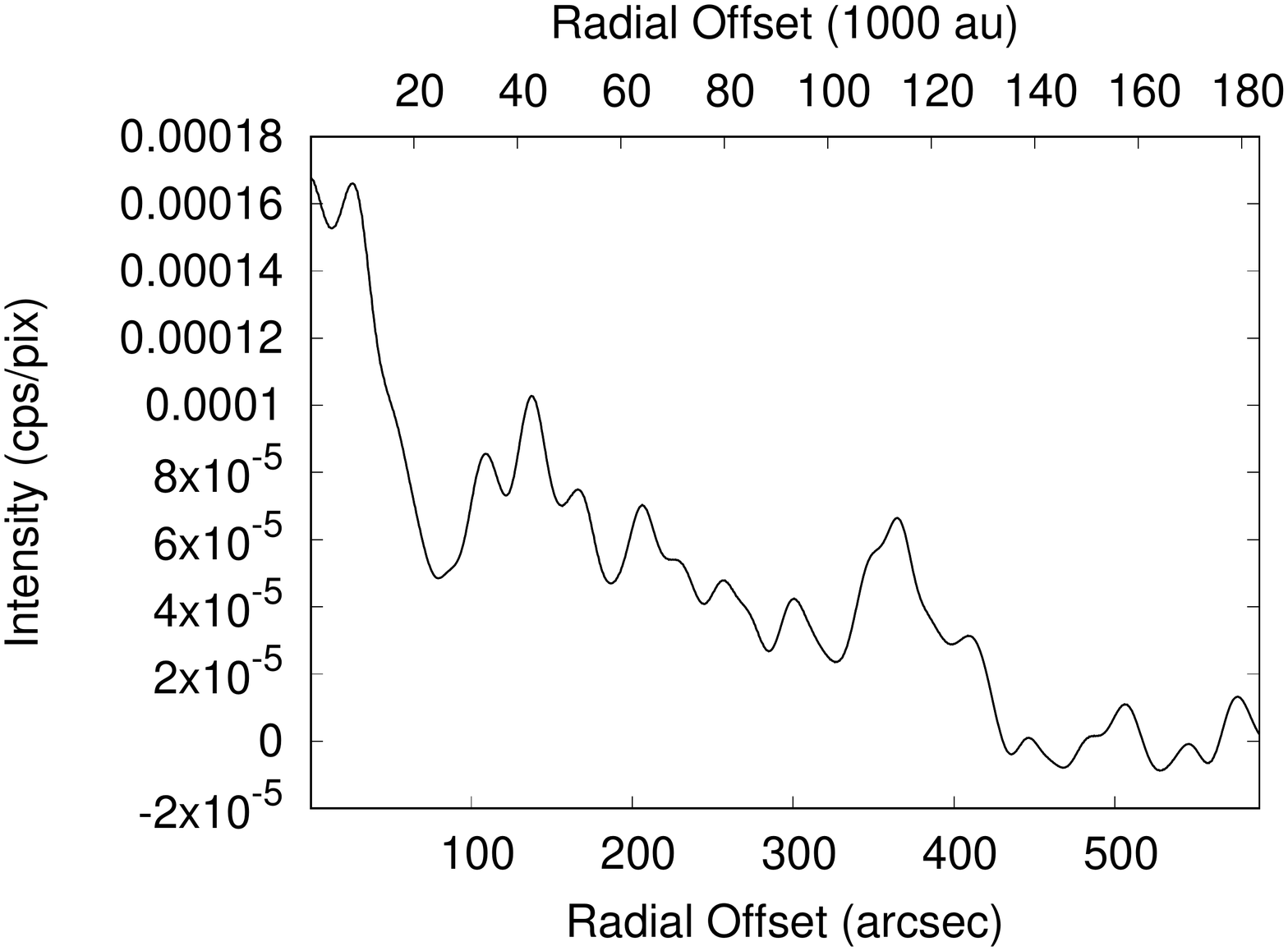}
\end{center}
\vspace{1.0in}
\caption{Radial intensity cut of the FUV emission around V\,Hya, averaged over an azimuthal wedge with its apex centered on the star, and spanning the range from PA=-100\arcdeg~to -80\arcdeg. The average  surrounding sky intensity has been subtracted from the cut.}
\label{VHyaCuts}
\end{figure}

\clearpage

\clearpage

\clearpage

\clearpage

\clearpage

\clearpage

\begin{figure}[htbp]
\begin{center}
\includegraphics[width=16cm]{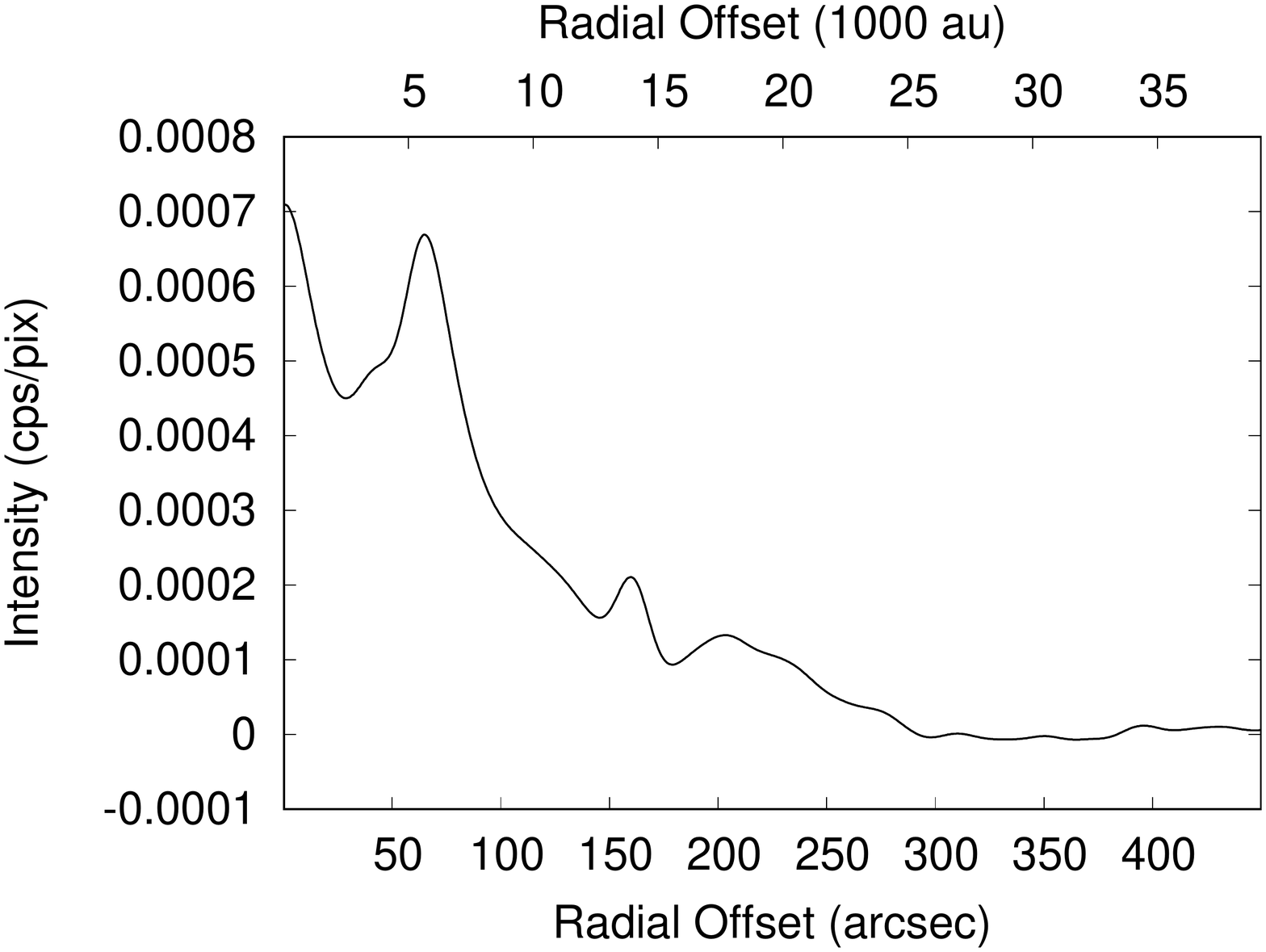}
\end{center}
\vspace{1.0in}
\caption{Radial intensity cut of the FUV emission around W\,Hya; intensity has been averaged over an azimuthal wedge with its apex centered on the star and covering the angular range  $PA=95\arcdeg$ to $235\arcdeg$.  The average surrounding sky intensity has been subtracted from the cut.}
\label{WHyaCuts}
\end{figure}

\clearpage

\clearpage

\begin{figure}[htbp]
\begin{center}
\includegraphics[width=16cm]{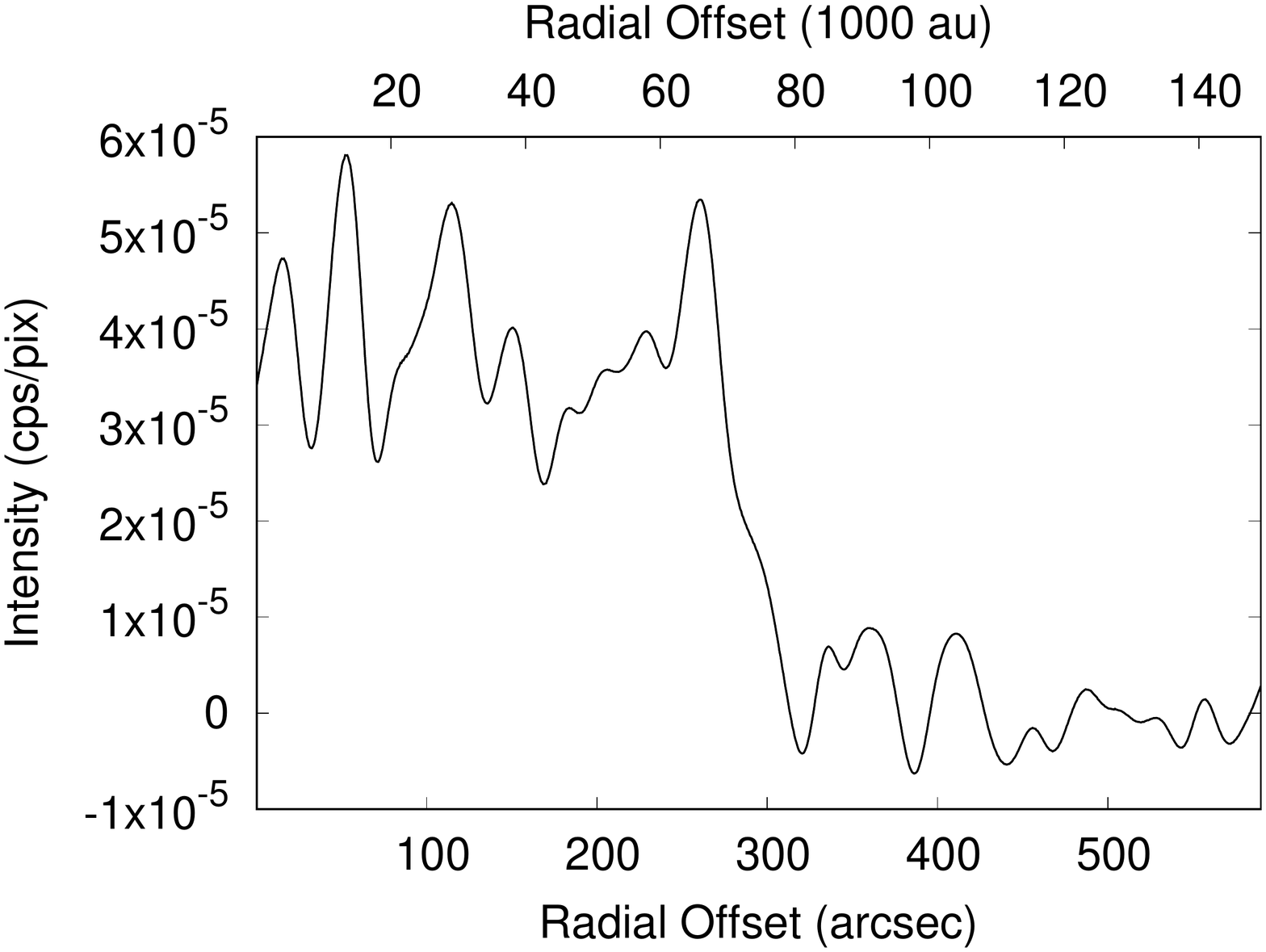}
\end{center}
\vspace{1.0in}
\caption{Radial intensity cut of the FUV emission around RW\,Boo; intensity has been averaged over an azimuthal wedges with its apex centered on the star, covering the angular range $PA=-75\arcdeg$ to $5\arcdeg$.  The average surrounding sky intensity has been subtracted from the cut.}
\label{RWBooCuts}
\end{figure}

\clearpage

\clearpage

\clearpage

\clearpage

\begin{figure}[htbp]
\begin{center}
\includegraphics[width=16cm]{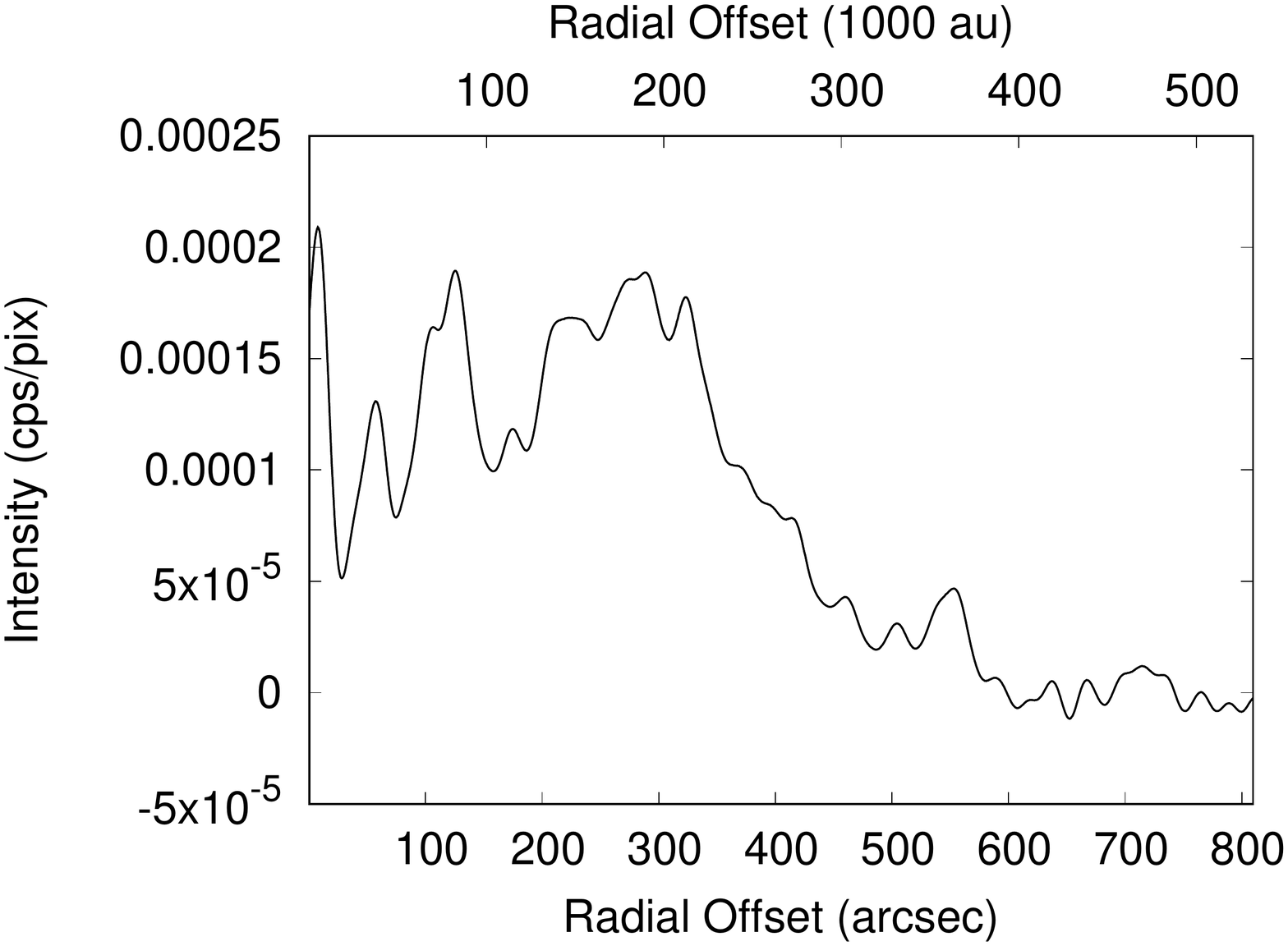}
\end{center}
\vspace{1.0in}
\caption{Radial intensity cut of the FUV emission around VX\,Eri; intensity has been averaged over an azimuthal wedge with its apex centered on the star and covering the angular range  $PA=100\arcdeg$ to $180\arcdeg$. The average surrounding sky intensity has been subtracted from the cut.}
\label{VXEriCuts}
\end{figure}

\clearpage

\clearpage

\clearpage
\appendix
The following table (Table\,\ref{tbl:allstars700s}) lists all AGB stars for which GALEX images were examined in order to search for extended UV emission associated with the stars, but where no such emission was found. The root names of the image fits files for these sources, together with the exposure times, Galactic coordinates, distance, height above the Galactic plane, and FUV extinction, are listed.
\clearpage
\voffset=-2in
\headsep=3in
\begin{longtable}{p{1.8cm} p{1.7cm} p{4.7cm} p{1.6cm} p{1.6cm} p{1.6cm} p{0.8cm} p{1.0cm} p{1.0cm}}
\caption{AGB Stars without Extended UV Emission}\\
Name     & Sp.Typ. & Image\,Root\,Name & Exp.Time & Long. & Lat.  & Dist. & $z$   & $A_{FUV}$ \\
(Simbad) &         &                   & (s)      & (Deg.)& (Deg.)& (kpc) & (kpc) &  \\
\hline
\endhead
TT Peg &  M3 & GI4\_042001\_AOHI000702p270035 & 1613  & 110.7468 & -34.7159 & 0.716 & -0.408 & 0.28   \\
AG Cet &  M5III & GI1\_026004\_Arp100 & 1593.05  & 102.2252 & -73.5815 & 0.279 & -0.268 & 0.14   \\
TU And &  M6e & GI1\_023001\_HIP2546 & 1611  & 117.5941 & -36.6447 & 1.010 & -0.603 & 0.26   \\
TW Psc &  M8 & MISDR1\_16795\_0418 & 1658  & 118.6546 & -48.5745 & 0.713 & -0.535 & 0.20   \\
57 Psc &  M4IIIa & MISDR1\_16810\_0419 & 1648  & 121.1921 & -47.3799 & 0.220 & -0.162 & 0.32   \\
CR Cet &  M4III & MISWZS01\_29157\_0269 & 2415.35  & 128.2863 & -64.4344 & 0.329 & -0.297 & 0.31   \\
Z Psc & C & GI1\_023021\_ZPsc & 3368.4  & 129.8623 & -36.7705 & 0.622 & -0.372 & 0.52   \\
AA Tri &  M3 & NGA\_NGC0777 & 1971.9  & 139.3226 & -29.6082 & 0.199 & -0.098 & 0.29   \\
R Cet &  M4$-$5e & GI1\_037003\_J022604p002135 & 6019.8  & 166.9656 & -54.7512 & 0.633 & -0.517 & 0.36   \\
X For &  M3 & MIS2DFSGP\_28387\_0161 & 1658.1  & 217.0063 & -65.1053 & 0.964 & -0.875 & 0.16   \\
RR Eri &  M5III & MISDR1\_18658\_0457 & 1696  & 185.1699 & -55.7683 & 0.375 & -0.310 & 0.35   \\
X Cet &  M5.5e & MISGCSS\_18775\_0410 & 3375.4  & 182.9203 & -45.9836 & 1.517 & -1.091 & 0.42   \\
GL Eri &  M5III & GI1\_047021\_ESO302\_G014 & 1696  & 242.3073 & -51.0352 & 0.309 & -0.240 & 0.18   \\
V Eri & M5/M6IV & GI1\_023013\_HIP19004 & 1616  & 208.8489 & -43.9775 & 0.297 & -0.207 & 0.19   \\
BD$-$05 836 &  M5 & MISDR1\_26911\_0465 & 3157.95  & 196.5405 & -37.9160 & 0.668 & -0.410 & 0.51   \\
V Cam & M7 & GI1\_023023\_VCam & 2300  & 139.3941 & 22.8971 & 0.627 & 0.244 & 1.13   \\
HD 45819 &  M5III & GI1\_099014\_NGC2249 & 1531.25  & 278.8037 & -27.6048 & 0.705 & -0.327 & 0.35   \\
AA Cam &  M5S & GI1\_023002\_HIP35045 & 1693  & 146.7974 & 27.2566 & 0.475 & 0.218 & 0.91   \\
VX Aur &  M4 & GI4\_016002\_DDO43 & 1688  & 177.6081 & 24.0286 & 1.018 & 0.414 & 1.13   \\
EY Cam &  M5 & GI3\_061009\_UGCA133 & 1693.05  & 148.4479 & 29.1185 & 0.519 & 0.253 & 0.89   \\
Y Gem &  M8 & MISGCSAN\_15326\_2078 & 1531.05  & 199.4917 & 19.7929 & 0.644 & 0.218 & 1.23   \\
SV Lyn &  M5III: & MISGCSAN\_04353\_0757 & 1522.1  & 184.5586 & 29.5006 & 0.244 & 0.120 & 0.50   \\
RZ UMa &  M8 & GI1\_023003\_HIP40060 & 1616  & 150.8722 & 32.7450 & 0.509 & 0.275 & 0.81   \\
RX Cnc &  M8 & MISGCSAN\_15612\_1585 & 1701.05  & 198.0768 & 28.5631 & 0.522 & 0.250 & 0.90   \\
RY Hya & Ce & MISGCSAN\_16455\_1184 & 1703  & 220.8651 & 20.9410 & 1.469 & 0.525 & 0.65   \\
Z Cnc &  M6III & GI4\_042008\_AOHI082329p150918 & 1639.05  & 209.0836 & 26.7663 & 0.419 & 0.189 & 0.44   \\
FW Cnc &  M0 & MISGCSAN\_16578\_1760 & 2931.25  & 216.8179 & 30.7204 & 0.365 & 0.187 & 0.38   \\
S Hya &  M4$-$M6.5e & MISWZN09\_24230\_0565 & 3156.05  & 224.9863 & 28.4174 & 1.116 & 0.531 & 0.49   \\
FZ Cnc &  M4IIIv & GI1\_113005\_SY\_CNC & 4794.35  & 209.5405 & 36.1078 & 0.221 & 0.130 & 0.24   \\
HD 77938 &  M4/M5III & GI2\_023004\_T\_PYX & 880  & 257.2315 & 9.6299 & 0.327 & 0.055 & 0.45   \\
NR Hya &  M8 & MISDR1\_24351\_0470 & 3174.05  & 230.8246 & 29.0731 & 0.662 & 0.322 & 0.47   \\
CW Cnc & M6 & MISGCSAN\_23934\_2434 & 1682.1  & 216.2250 & 36.2926 & 0.252 & 0.149 & 0.29   \\
IN Hya &  M3 & MISWZN09\_24315\_0213o & 1686.05  & 231.9017 & 32.7302 & 0.327 & 0.177 & 0.34   \\
DF Leo &  M4III & MISDR3\_24067\_1195 & 1689.05  & 224.4304 & 37.1814 & 0.327 & 0.197 & 0.32   \\
TW Sex &  M4 & MISDR1\_24339\_0267 & 3041.15  & 238.3158 & 39.6127 & 0.452 & 0.288 & 0.35   \\
UY Leo &  M7III: & GI1\_047040\_UGC05672 & 1766.1  & 212.1644 & 57.6836 & 0.715 & 0.605 & 0.26   \\
S Sex &  M4$-$5e & MISDR1\_24330\_0273 & 1690.1  & 247.2382 & 47.2226 & 1.353 & 0.993 & 0.30   \\
GV UMa &  M5 & LOCK\_08 & 12087.3  & 152.2240 & 52.5242 & 0.568 & 0.451 & 0.55   \\
R UMa &  M5$-$8e & GI1\_023015\_HIP52546 & 1703  & 138.3629 & 44.3614 & 0.560 & 0.391 & 0.64   \\
VY UMa & C & GI1\_023025\_VYUma & 1704  & 139.5897 & 45.4137 & 0.416 & 0.296 & 0.61   \\
GY UMa &  M4III & GI4\_016006\_DDO87 & 1533.05  & 141.0463 & 46.9199 & 0.307 & 0.225 & 0.55   \\
56 Leo &  M5.5III & MISWZN11\_12423\_0315 & 1533.05  & 245.0495 & 55.4968 & 0.115 & 0.095 & 0.09   \\
FF Leo &  M5 & MISWZN11\_12490\_0315 & 954.55  & 249.4775 & 55.3027 & 0.858 & 0.705 & 0.27   \\
AK Leo &  M... & GI5\_039002\_AGC215158\_HA1 & 1585.15  & 249.8324 & 68.3165 & 0.648 & 0.602 & 0.23   \\
R Com &  M5$-$7e & GI1\_079001\_NGC4064 & 1691.05  & 248.0329 & 76.3160 & 1.261 & 1.225 & 0.22   \\
FZ Vir &  M... & MISDR1\_13708\_0334 & 1513.5  & 291.7645 & 59.9583 & 0.402 & 0.348 & 0.25   \\
T CVn &  M5 & GI4\_015003\_DDO133 & 9973.5  & 168.2676 & 83.6332 & 0.800 & 0.795 & 0.43   \\
BZ Vir &  M5 & GI1\_047084\_UGCA319 & 1462  & 306.0581 & 45.1690 & 2.486 & 1.763 & 0.32   \\
SY CVn &  M8 & GI1\_047086\_UGC08215 & 1572.2  & 113.6014 & 69.7432 & 0.760 & 0.713 & 0.23   \\
FH Vir &  M6III & GI1\_023005\_HIP64768 & 1665.1  & 320.1010 & 68.5401 & 0.359 & 0.334 & 0.23   \\
RW CVn &  M7III: & GI1\_026018\_Arp84 & 2811.4  & 72.2044 & 72.4610 & 0.462 & 0.441 & 0.23   \\
BY Boo &  M4.5:III & MISGCSN\_01277\_1394 & 1222.1  & 85.2586 & 67.2613 & 0.166 & 0.153 & 0.15   \\
FS Vir &  M4III & MISDR1\_33714\_0583 & 1690.1  & 346.5112 & 58.9557 & 0.248 & 0.213 & 0.21   \\
NO Vir &  M5 & MISWZN15\_33932\_0237 & 1575.95  & 342.8404 & 53.3535 & 0.497 & 0.399 & 0.28   \\
AO Vir &  M4 & MISDR1\_33712\_0584 & 2504.1  & 349.7972 & 58.2956 & 3.514 & 2.990 & 0.26   \\
S Boo &  M5$-$6e & PS\_GROTH\_MOS01 & 4048.35  & 96.9253 & 58.4554 & 1.595 & 1.359 & 0.26   \\
RS Vir &  M8 & MISWZN15\_33657\_0360o & 1630  & 352.6741 & 57.9715 & 0.453 & 0.384 & 0.26   \\
NV Vir &  M5III & MISWZN15\_33963\_0340 & 1942.1  & 348.2470 & 49.3709 & 0.773 & 0.587 & 0.30   \\
NU Aps &  M5III & GI4\_099003\_IC4499 & 4295.9  & 306.8874 & -20.7643 & 0.600 & -0.213 & 0.46   \\
Y Ser &  M5e & MISWZN15\_33920\_0338 & 2493.2  & 358.4771 & 45.0585 & 0.436 & 0.309 & 0.32   \\
Z Ser &  M5 & MISDR1\_33757\_0591 & 1668.05  & 3.3780 & 47.3378 & 0.907 & 0.667 & 0.31   \\
Y CrB &  M8III: & GI1\_023007\_HIP77284 & 2883.1  & 61.3352 & 51.8605 & 0.686 & 0.539 & 0.29   \\
X Her &  M8 & GI5\_021006\_X\_Her & 3024.1  & 74.4645 & 47.7858 & 0.123 & 0.091 & 0.09   \\
FQ Ser &  M3 & MISGCSAN\_22074\_1730 & 1645.05  & 20.8265 & 39.9553 & 0.180 & 0.116 & 0.18   \\
RU Her &  M6$-$7e & GI1\_023008\_HIP79233 & 2426.7  & 41.9779 & 45.6111 & 0.684 & 0.489 & 0.32   \\
TV Dra &  MS & MISDR1\_09998\_0349 & 2584  & 94.3435 & 35.3483 & 0.510 & 0.295 & 0.39   \\
V945 Her &  M5 & MISDR2\_22240\_0978 & 2717.05  & 53.6682 & 32.1726 & 1.372 & 0.731 & 0.45   \\
T Dra & Nev & GI1\_023020\_HIP87820 & 1612.45  & 86.7498 & 29.9422 & 0.944 & 0.471 & 0.47   \\
SS Lyr &  M5IIIe & GI4\_056015\_KEPLER\_06 & 5800.05  & 77.9802 & 16.0017 & 0.748 & 0.206 & 0.81   \\
GY Aql &  M8 & GI1\_023009\_HIP97586 & 2608.25  & 32.7232 & -16.4858 & 0.706 & -0.200 & 0.59   \\
UX PsA &  M3/M4III & MIS2DFSGP\_40469\_0326 & 1527.1  & 16.7881 & -48.0706 & 0.397 & -0.295 & 0.20   \\
TU Peg &  M7$-$8e & MISDR2\_20095\_0732 & 1937.2  & 68.2077 & -29.7630 & 0.620 & -0.308 & 0.33   \\
EP Aqr &  M8IIIv & MISWZS22\_20562\_0261 & 2759  & 54.2001 & -39.2604 & 0.133 & -0.084 & 0.08   \\
KL Aqr &  M8 & MISGCSN\_20728\_0261o & 2279.3  & 59.7637 & -41.2647 & 0.505 & -0.333 & 0.24   \\
SV Peg & M7 & GI1\_023010\_HIP109070 & 1724.05  & 88.7152 & -16.2856 & 0.401 & -0.113 & 0.44   \\
TX Peg &  M... & MISDR2\_20490\_0736 & 1687.15  & 75.7147 & -34.8694 & 0.550 & -0.314 & 0.28   \\
S Gru &  M8IIIe & GI1\_047110\_ESO238\_G005 & 2415.2  & 345.8846 & -54.7673 & 0.561 & -0.458 & 0.18   \\
AF Peg & M5II$-$III & GI1\_023011\_HIP112868 & 1646.9  & 86.8368 & -36.2020 & 0.402 & -0.238 & 0.26   \\
TY And & M... & GI1\_023012\_HIP114757 & 1887  & 103.8529 & -18.4676 & 0.662 & -0.210 & 0.50   \\
SV Aqr & M8 & MISDR1\_29581\_0645 & 2733.05  & 66.6906 & -63.5235 & 0.422 & -0.378 & 0.16   \\
LW Aqr &  M4III & MISWZS00\_29619\_0271 & 1690.25  & 77.7709 & -69.5664 & 0.405 & -0.379 & 0.15   \\
XZ Psc &  M5III & MISWZS00\_29073\_0165 & 4289.4  & 94.0594 & -59.5482 & 0.180 & -0.155 & 0.12   \\

\label{tbl:allstars700s}
\end{longtable}

\end{document}